\documentclass[11pt]{article}

\usepackage[a4paper,margin=1in]{geometry}
\usepackage{amsmath,amssymb,amsthm,mathtools}
\usepackage{booktabs}
\usepackage{array}
\usepackage{graphicx}
\usepackage{enumitem}
\usepackage{cite}
\usepackage{hyperref}
\usepackage{xcolor}

\hypersetup{
 colorlinks=true,
 linkcolor=blue,
 citecolor=blue,
 urlcolor=blue
}

\newtheorem{theorem}{Theorem}
\newtheorem{lemma}{Lemma}
\newtheorem{proposition}{Proposition}
\newtheorem{corollary}{Corollary}
\newtheorem{remark}{Remark}
\newtheorem{definition}{Definition}
\newtheorem{example}{Example}

\newcommand{\F}{\mathbb F}
\newcommand{\N}{\mathcal N}
\newcommand{\C}{\mathcal C}
\newcommand{\supp}{\operatorname{supp}}
\newcommand{\wtH}{\operatorname{wt}_{H}}
\newcommand{\wtL}{\operatorname{wt}_{L}}
\newcommand{\Span}{\operatorname{Span}}

\begin{document}
\title{\textbf{Four Classes of Few-Weight Linear Codes over 
\\ $u\mathbb{F}_q+u^2\mathbb{F}_q+u^3\mathbb{F}_q$ from Simplicial Complexes}}

\author{
Lan Deng\textsuperscript{1,2},
Zhao Hu\textsuperscript{3},
Nian Li\textsuperscript{3}
}

\date{}
\maketitle

\begingroup
\renewcommand{\thefootnote}{}
\footnotetext{Email: denglan2026@126.com, zhao.hu@aliyun.com, nian.li@hubu.edu.cn.}
\endgroup

\begingroup
\setcounter{footnote}{0}
\renewcommand{\thefootnote}{\arabic{footnote}}
\footnotetext[1]{Hubei Key Laboratory of Applied Mathematics, Faculty of Mathematics and Statistics, Hubei University, Wuhan 430062, China.}

\footnotetext[2]{School of Mathematics and Statistics, Hanjiang Normal University, Shiyan 442000, China.}

\footnotetext[3]{Hubei Provincial Engineering Research Center of Intelligent Connected Vehicle Network Security, School of Cyber Science and Technology, Hubei University, Wuhan 430062, China.}
\setcounter{footnote}{3}
\endgroup

\begin{abstract}
In this paper, we construct linear codes over the commutative non-unital ring $u\mathbb{F}_q+u^2\mathbb{F}_q+u^3\mathbb{F}_q$, where $u^4=0$, using simplicial complexes in $\mathbb{F}_q^m$ and
their complements. For codes defined by two arbitrary nonempty simplicial complexes, we determine the lengths and $\mathbb{F}_q$-dimensions. When each complex has a single maximal element with nonempty proper support, we use an $\mathbb{F}_q$-linear Gray map and exponential sums to determine the parameters and Lee weight distributions of four classes of codes. These classes have two, at most five, at most four, and at most eight nonzero Lee weights, respectively. Puncturing the Gray images yields projective codes with dual minimum distance three. Using the Griesmer bound, we obtain families of distance-optimal projective codes, including a binary family whose lengths exceed the bound by three. The Gray images are Euclidean self-orthogonal for $q=3$ and for $q=2$ under an additional condition, and Hermitian self-orthogonal for $q=4$. We also give sufficient conditions for minimality of both the Gray images and the projective codes.
\end{abstract}

\noindent\textbf{Keywords:}
non-unital ring; simplicial complex; defining set;
Gray map; Lee weight; few-weight code.

\noindent\textbf{Mathematics Subject Classification:} 11T71, 11T23, 14G50

\section{Introduction}

Let $\F_q$ be the finite field with $q$ elements, where $q$ is a power of a prime $p$. An $[n,k,d]$ linear code $\C$ over $\F_q$ is a $k$-dimensional subspace of $\F_q^n$ with minimum Hamming distance $d$. For $0\leq w\leq n$, let $A_w$ denote the number of codewords of Hamming weight $w$ in $\C$. The polynomial $1+A_1z+A_2z^2+\cdots+A_nz^n$ is the weight enumerator of $\C$, and the sequence $(1,A_1,A_2,\ldots,A_n)$ is called its weight distribution. The weight distribution gives information about the error-correcting capability of a code and is used in the analysis of error detection and correction~\cite{HuffmanPless,Klove2007}. A code is called a $t$-weight code if exactly $t$ entries in $(A_1,A_2,\ldots,A_n)$ are nonzero. Linear codes with few weights have applications in secret sharing schemes~\cite{ADHK, CarletDingYuan2005}, authentication codes~\cite{DingHellesethKloveWang2007}, and strongly regular graphs~\cite{CalderbankKantor1986}. An $[n,k,d]$ linear code over $\F_q$ is said to be distance-optimal if no $[n,k,d+1]$ linear code over $\F_q$ exists. The Griesmer bound~\cite{Griesmer1960} gives a lower bound on the length of a linear code in terms of its dimension and minimum distance. A code is called a Griesmer code if its length attains this bound, and a near Griesmer code if its length exceeds the bound by one.

In 2007, Ding and Niederreiter~\cite{DingNiederreiter2007} introduced a general construction of linear codes using trace functions and defining sets. In the vector-space form of this construction, a codeword
is obtained by taking the inner products of a parameter vector with the elements of an ordered defining set in $\F_q^m$. The choice of the defining set determines the parameters and weight distribution of the resulting code. In particular, the Hamming weight of a codeword can be calculated by counting the elements of the defining set for which the corresponding inner product is zero. These counts can be expressed as exponential sums using additive-character orthogonality. By choosing suitable defining sets, many linear codes with good parameters have been constructed; see, for example, \cite{HengDingWang2020,Heng2023,HyunLeeLee2020,LiYueTang2022,MesnagerQiRuTang2020,MesnagerSinak2020,ZhouLiFanHelleseth2016} and the references therein.

Simplicial complexes provide a class of defining sets whose character sums can be evaluated by combining the orthogonality of additive characters with inclusion-exclusion, since they are unions of coordinate subspaces. In 2018, Chang and Hyun~\cite{ChangHyun2018} used simplicial complexes of $\F_2^m$ to construct binary linear codes. Subsequent studies have produced various classes of few-weight linear codes over finite fields and finite rings from simplicial complexes \cite{BhagatSarmaSagar2025,HyunLeeLee2020,LiShiBinary2021,LiShiTrace2022,LiuYu2024,MondalLee2024,SagarSarmaOctanary2023,SagarSarma2024TIT,SagarSarma2024AMC,ShiLi2022,WuLee2020,WuLiFanXiao2024,WuLiXiao2022,WuLiZhangXiao2024,WuZhuYue2020,YadavSarmaBhagat2025,ZhuWei2021}. Several of these constructions yield infinite families of optimal linear codes. For binary codes, Hyun, Lee, and Lee~\cite{HyunLeeLee2020} constructed infinite families of optimal linear codes from general simplicial complexes, without restricting the number of maximal elements. Over prime fields, constructions from down-sets were studied by Hyun, Kim, and Na~\cite{HyunKimNa2019} and Shi and Li~\cite{ShiLi2021}. Hu et al.~\cite{HuXuEtAl2024} constructed infinite families of optimal linear codes over arbitrary finite fields from general simplicial complexes. More recently, Hu et al.~\cite{HuEtAl2026} used a bivariate defining-set construction with simplicial complexes having one or two maximal elements. They determined the parameters and weight distributions of several classes of few-weight linear codes and investigated their optimality and minimality. For further developments, see the survey by Wu, Li, and Hyun~\cite{WuLiHyunSurvey2024} and the references therein. Over the ring $\F_2+u\F_2$ with $u^2=0$, Wu, Zhu, and Yue~\cite{WuZhuYue2020} constructed few-Lee-weight codes using the standard inner product and simplicial complexes with one maximal element. Their Gray images include binary Griesmer codes and distance-optimal codes. Shi and Li~\cite{ShiLi2022} studied few-weight codes over a non-chain ring associated with simplicial
complexes and investigated the optimality of their Gray images. Chen et al.~\cite{ChenEtAl2025} constructed four families of trace codes over $\F_q+u\F_q$, where $u^2=0$, using differences of simplicial complexes and their complements. They determined the parameters and Lee weight distributions and studied the optimality of the Gray images. Yadav, Tiwari, and Sarma~\cite{YadavTiwariSarma2026} also studied codes over $\F_q+u\F_q$ constructed from simplicial complexes, together with their Gray images and subfield codes. For non-unital rings, Sagar and
Sarma~\cite{SagarSarma2024TIT} constructed codes over a non-commutative ring of order four using simplicial complexes with one maximal element. They also studied codes over another non-unital ring
in~\cite{SagarSarma2024DCC}. Wu et al.~\cite{WuEtAl2026} considered extensions of non-unital non-commutative rings over arbitrary finite fields. They determined the principal parameters for defining sets formed from general simplicial complexes and obtained explicit Lee weight distributions when each complex has one maximal element. They also investigated self-orthogonality, minimality, optimality, and projectiveness through the Gray images and subfield-like codes. In the commutative non-unital setting, Sagar, Patel, and Singh~\cite{SagarPatelSingh2026} constructed codes from general simplicial complexes and studied properties of their Gray images.

Motivated by these works, we construct linear codes $\C_D$ over the commutative non-unital ring $\N_q=u\F_q+u^2\F_q+u^3\F_q$ of order $q^3$, where $u^4=0$, from simplicial complexes of $\F_q^m$ and their complements via the defining-set approach. We introduce an $\F_q$-linear Gray map $\phi$ from $\N_q$ to $\F_q^3$. For $D_1,D_2,D_3\subseteq\F_q^m$ with $D_3\ne\varnothing$, the code defined by $uD_1+u^2D_2+u^3D_3$ is obtained, up to a coordinate permutation, by repeating each coordinate of the code defined by $uD_1+u^2D_2$ exactly $|D_3|$ times. When $D_1$ and $D_2$ are arbitrary nonempty simplicial complexes, we determine the length and $\F_q$-dimension of the code defined by $uD_1+u^2D_2$. We derive a general Lee-weight formula by additive-character orthogonality and use inclusion-exclusion to evaluate the character sums associated with simplicial complexes. When the first two complexes each have a single maximal element, with supports $A$ and $B$, respectively, where $\varnothing\neq A,B\subsetneq[m]$, we consider the four defining sets $D=u\Delta_A+u^2\Delta_B$ (Case~I), $D=u\Delta_A^c+u^2\Delta_B$ (Case~II), $D=u\Delta_A+u^2\Delta_B^c$ (Case~III), and $D=u\Delta_A^c+u^2\Delta_B^c$ (Case~IV), with complements taken in $\F_q^m$. We determine the parameters and Lee weight distributions of the resulting four classes, which have two, at most five, at most four, and at most eight nonzero Lee weights, respectively. We also give necessary and sufficient conditions for the last three bounds to be attained. Puncturing the Gray images gives projective codes $\overline{\C}$ with dual minimum distance three. We determine their lengths and dimensions in all four cases and their weight distributions in Case~III and in Case~IV with $B\subseteq A$. Using the Griesmer bound, we obtain families of distance-optimal codes in Case~III. The binary examples with parameters $[13,4,6]$ from Case~I and $[10,4,4]$ and $[36,6,16]$ from Case~IV are also distance-optimal. The last two examples occur in Case~II as well. Furthermore, the Gray images $\phi(\C_D)$ of these four classes are Euclidean self-orthogonal for $q=3$ and for $q=2$ when $|A|+|B|\geq 3$, and Hermitian self-orthogonal for $q=4$. Finally, we apply the Ashikhmin--Barg criterion to obtain sufficient conditions for both $\phi(\C_D)$ and $\overline{\C}$ to be minimal.  Several examples verified by Magma computations are provided to illustrate the results.

The rest of this paper is organized as follows. Section~\ref{sec:preliminaries} recalls the notation, simplicial complexes, the Gray map, and the auxiliary results used in the paper. Section~\ref{sec:construction} presents the defining-set construction and determines the parameters and Lee weight distributions of the four classes of codes. Section~\ref{sec:applications} constructs projective codes by puncturing the Gray images and studies their distance optimality. It then discusses self-orthogonality and minimality. Section~\ref{sec:conclusion} concludes the paper.

\section{Preliminaries}\label{sec:preliminaries}

In this section, we recall the notation, definitions, and auxiliary results used in the subsequent sections. Throughout this paper, the following notation will be used unless otherwise stated.
\begin{itemize}
\item Let $q$ be a power of a prime $p$, and let $m$ be a positive integer. Denote the finite field with $q$ elements by $\F_q$, and put $\F_q^*=\F_q\setminus\{0\}$.
\item For a positive integer $l$, write $[l]=\{1,2,\ldots,l\}$.
\item For a finite set $E$, denote its cardinality by $|E|$. For $\Delta\subseteq\F_q^m$, denote its complement in $\F_q^m$ by $\Delta^c=\F_q^m\setminus\Delta$.
\item For a vector $x$ over $\F_q$, let $\wtH(x)$ denote its Hamming weight. For vectors $x,y$ of the same length, write $d_H(x,y)=\wtH(x-y)$ for their Hamming distance.
\item For $x=(x_1,\ldots,x_m)\in\F_q^m$, let $\supp(x)=\{i\in[m]:x_i\neq0\}$ be its support.
\item For $x,y\in\F_q^m$, write $\langle x,y\rangle_q=\sum_{i=1}^m x_i y_i$ for their inner product over $\F_q$.
\item Let $\operatorname{Tr}_p^q$ denote the trace from $\F_q$ to $\F_p$, and let $\zeta_p$ be a primitive complex $p$-th root of unity. The canonical additive character of $\F_q$ is denoted by $\chi$, where $\chi(x)=\zeta_p^{\operatorname{Tr}_p^q(x)}$.
\item For a simplicial complex, $\mathcal F$ denotes its set of maximal elements with entries $0$ or $1$, and $\mathcal A$ denotes the family of their supports. The letters $A,B,F_i,H_j$ denote support sets. The symbols $\Delta_1,\Delta_2$ denote general simplicial complexes. When they have one maximal element each, we write $\Delta_1=\Delta_A$ and $\Delta_2=\Delta_B$.
\item The symbol $A_w$ denotes the number of codewords of Hamming weight $w$ for a code over $\F_q$, or of Lee weight $w$ for a code over $\N_q$, as appropriate.
\item Denote the ring defining set by $D$, its component sets in $\F_q^m$ by $D_1,D_2,D_3$, and the corresponding code by $\C_D$. Denote the Gray map by $\phi$ and the Lee weight by $\wtL$.
\item For a linear code $\C$ over $\F_q$, denote its Euclidean dual by $\C^\perp$; for $q=4$, denote its Hermitian dual by $\C^{\perp_H}$. The projective code obtained from $\phi(\C_D)$ by deleting zero columns and retaining one column from each class of proportional nonzero columns is denoted by $\overline\C$, and its length by $\overline n$.
\end{itemize}

\subsection{Simplicial complexes of \texorpdfstring{$\F_q^m$}{Fq m}}\label{subsec:simplicial}

The following definitions and properties can be found in~\cite[Section~2.1]{HuEtAl2026} and~\cite[Section~II-C]{WuEtAl2026}. For two vectors $x,y\in\F_q^m$, we say that $x$ covers $y$, denoted by $y\preceq x$, if $\supp(y)\subseteq\supp(x)$. A subset $\Delta\subseteq\F_q^m$ is called a simplicial complex if $x\in\Delta$ and $y\preceq x$ imply $y\in\Delta$. Every nonempty simplicial complex contains the zero vector.

An element $x\in\Delta$ with entries $0$ or $1$ is called maximal if no element of $\Delta$ has support properly containing $\supp(x)$. Let $\mathcal F$ be the set of maximal elements of a nonempty complex $\Delta$. The family $\mathcal A=\{\supp(x):x\in\mathcal F\}$ is called the support of $\Delta$. A simplicial complex is uniquely determined by this family, and we write $\Delta=\langle\mathcal A\rangle$. For $F\subseteq[m]$, write $\Delta_F=\{x\in\F_q^m:\supp(x)\subseteq F\}$. This is the simplicial complex with one maximal element of support $F$. A general simplicial complex is the union of the coordinate subspaces indexed by the members of its support: $\Delta=\langle\mathcal A\rangle=\bigcup_{F\in\mathcal A}\Delta_F$.
In particular, $\Delta_A=\langle\{A\}\rangle$ is an $|A|$-dimensional $\F_q$-subspace of $\F_q^m$, so that $|\Delta_A|=q^{|A|}$ and $|\Delta_A^c|=q^m-q^{|A|}$. If $A=\varnothing$, then $\Delta_A=\{0\}$. For $A,B\subseteq[m]$, we also have $\Delta_A\cap\Delta_B=\Delta_{A\cap B}$ and $\Delta_A+\Delta_B=\Delta_{A\cup B}$. Since scalar multiplication by $\F_q^*$ preserves supports, every simplicial complex and its complement are stable under multiplication by $\F_q^*$.

\subsection{Lee weight and Gray map}\label{subsec:gray}

Consider the ring $\F_q[u]/\langle u^4\rangle$ and its maximal ideal
\[
\N_q=u\F_q+u^2\F_q+u^3\F_q.
\]
The addition and multiplication in $\N_q$ are inherited from the quotient ring. Every element
of $\N_q$ is uniquely written as $ux_1+u^2x_2+u^3x_3$, where $x_1,x_2,x_3\in\F_q$. Consequently, $\N_q$ is a commutative ring of order $q^3$, with $\F_q$-basis $\{u,u^2,u^3\}$. Since $u^4=0$, its multiplication is given by $(ux_1+u^2x_2+u^3x_3)(uy_1+u^2y_2+u^3y_3)=u^2x_1y_1+u^3(x_1y_2+x_2y_1)$ for $x_i,y_i\in\F_q$, $i=1,2,3$. The ring has no multiplicative identity, since $u\neq0$ and every product of $u$ with an element of $\N_q$ lies in $u^2\F_q+u^3\F_q$.

\begin{definition}[{\cite[Section~II-A]{WuEtAl2026}}]\label{def:ring-code}
An $\N_q$-submodule of $\N_q^n$ is called a linear code over $\N_q$ of length $n$.
\end{definition}

Thus a linear code over $\N_q$ is an additive subgroup of $\N_q^n$ closed under multiplication by elements of $\N_q$. Define the Gray map $\phi:\N_q\to\F_q^3$ by
\[
\phi(ux_1+u^2x_2+u^3x_3)
=(x_1+x_2,x_1+x_3,x_1+x_2+x_3).
\]
The map is extended coordinatewise to $\N_q^n$: for $x=(x_1,\ldots,x_n)\in\N_q^n$, let
\[
\phi(x)=(\phi(x_1),\ldots,\phi(x_n))\in\F_q^{3n}.
\]
The Lee weight of $x\in\N_q^n$ is defined by $\wtL(x)=\wtH(\phi(x))$, and the Lee distance between $x,y\in\N_q^n$ is $d_L(x,y)=\wtL(x-y)$.

\begin{proposition}\label{prop:gray}
The Gray map $\phi$ is an $\F_q$-linear bijection. Moreover,
\[
d_L(x,y)=d_H(\phi(x),\phi(y))
\]
for all $x,y\in\N_q^n$.
\end{proposition}

\begin{proof}
With respect to the basis $\{u,u^2,u^3\}$, the Gray map is represented by the matrix
\[
\begin{pmatrix}
1&1&0\\
1&0&1\\
1&1&1
\end{pmatrix},
\]
whose determinant is $-1$. Hence $\phi$ is an $\F_q$-linear bijection. Its coordinatewise extension is also linear and bijective. By the definition of the Lee weight and the linearity of $\phi$, we obtain $d_L(x,y)=\wtH(\phi(x-y))=\wtH(\phi(x)-\phi(y))=d_H(\phi(x),\phi(y))$. This completes the proof.
\end{proof}

For a linear code $\C$ over $\N_q$ of length $n$, let $A_w$ denote the number of codewords of Lee weight $w$. Its Lee weight enumerator is $1+A_1z+A_2z^2+\cdots+A_{3n}z^{3n}$, and the sequence $(1,A_1,A_2,\ldots,A_{3n})$ is its Lee weight distribution. A code is called a $t$-Lee-weight code if exactly $t$ entries in $(A_1,A_2,\ldots,A_{3n})$ are nonzero. For a nonzero linear code, the minimum Lee distance is the smallest Lee weight of a nonzero codeword. Proposition~\ref{prop:gray} shows that the Lee weight distribution and minimum Lee distance of $\C$ coincide with the Hamming weight distribution and minimum Hamming distance of $\phi(\C)$. The defining-set codes in Section~\ref{sec:construction} are $\F_q$-linear, and therefore their Gray images are linear codes over $\F_q$.

\subsection {Some useful definitions and lemmas} \label{subsec:useful-lemmas}

For a subset $E\subseteq\F_q^m$, its orthogonal complement is defined by $E^\perp=\{y\in\F_q^m:
\langle y,t\rangle_q=0\text{ for all }t\in E\}$. It is an $\F_q$-subspace of $\F_q^m$. If $E$ is an $r$-dimensional subspace, then $\dim_{\F_q}E^\perp=m-r$. For $A\subseteq[m]$, one has $\Delta_A^\perp=\Delta_{[m]\setminus A}=\{y\in\F_q^m:y_i=0\text{ for all }i\in A\}$, and hence $|\Delta_A^\perp|=q^{m-|A|}$.
By~\cite[Lemma~3]{WuEtAl2026}, $\Delta_A^\perp\cap\Delta_B^\perp=\Delta_{A\cup B}^\perp$, whose cardinality is $q^{m-|A\cup B|}$. The cosets $\Delta_A^\perp$ and $-y+\Delta_A^\perp$ coincide if $y\in\Delta_A^\perp$ and are disjoint otherwise.

The canonical additive character satisfies $\chi(x+y)=\chi(x)\chi(y)$ for $x,y\in\F_q$; see~\cite{LidlNiederreiter}. We use the following orthogonality identity over subspaces.

\begin{lemma}[{\cite[Lemma~4]{WuEtAl2026}}]\label{lem:orth-subspace}
Let $E$ be an $r$-dimensional $\F_q$-subspace of $\F_q^m$. Then, for $y\in\F_q^m$, we have
\[
\sum_{t\in E}\chi(\langle y,t\rangle_q)=
\begin{cases}
q^r,& \text{if }y\in E^\perp;\\
0,& \text{if }y\notin E^\perp.
\end{cases}
\]
\end{lemma}

For $E\subseteq\F_q^m$ and $y\in\F_q^m$, write $S_E(y)=\sum_{t\in E}\chi(\langle y,t\rangle_q)$. If $E$ is stable under multiplication by $\F_q^*$, then the substitution $t\mapsto\lambda t$ gives $S_E(\lambda y)=S_E(y)$ for every $\lambda\in\F_q^*$.

For a simplicial complex $\Delta$, scalar invariance gives $\sum_{\lambda\in\F_q}S_\Delta(\lambda y)
=|\Delta|+(q-1)S_\Delta(y)$. Thus the inclusion-exclusion identities take the following form.

\begin{lemma}[{\cite[Lemmas~1 and~5]{WuEtAl2026}}]\label{lem:simp-char}
Let $\Delta$ be a nonempty simplicial complex whose maximal elements have supports $\mathcal A=\{F_1,\ldots,F_l\}$, where $F_i\subseteq[m]$. For every $y\in\F_q^m$,
\[
S_\Delta(y)
=
\sum_{\substack{\varnothing\neq I\subseteq[l]\\
 y\in\Delta_{\cap_{i\in I}F_i}^{\perp}}}
(-1)^{|I|+1}q^{|\cap_{i\in I}F_i|}.
\]
Moreover,
\[
S_{\Delta^c}(y)
=
\begin{cases}
q^m-|\Delta|,& \text{if }y=0;\\
-S_\Delta(y),& \text{if }y\neq0.
\end{cases}
\]
In particular,
\[
|\Delta|=S_\Delta(0)
=\sum_{\varnothing\neq I\subseteq[l]}
(-1)^{|I|+1}q^{|\cap_{i\in I}F_i|}.
\]
\end{lemma}

To study the optimality of the codes constructed in this paper, we next recall the Griesmer bound.

\begin{lemma}[{\cite{Griesmer1960,SolomonStiffler1965}}]\label{lem:griesmer}
Let $\C$ be an $[n,k,d]$ linear code over $\F_q$. Then
\begin{equation}\label{eq:griesmer-bound}
n\ge g(k,d):=\sum_{i=0}^{k-1}\left\lceil\frac{d}{q^i}\right\rceil,
\end{equation}
where $\lceil\cdot\rceil$ denotes the ceiling function.
\end{lemma}

In the notation of Eq.~\eqref{eq:griesmer-bound}, a code is a Griesmer code if $n=g(k,d)$ and a near Griesmer code if $n-1=g(k,d)$. By Lemma~\ref{lem:griesmer}, an $[n,k,d]$ linear code is distance-optimal whenever $g(k,d+1)>n$, since an $[n,k,d+1]$ linear code would contradict Eq.~\eqref{eq:griesmer-bound}. These comparisons of length and Griesmer sums will be used in Section~\ref{sec:applications}.

We also recall a criterion for self-orthogonality. A linear code $\C\subseteq\F_q^n$ is Euclidean
self-orthogonal if $\C\subseteq\C^\perp$. For $q=4$, the Hermitian inner product is $\langle x,y\rangle_H=\sum_{i=1}^n x_i y_i^2$, and the corresponding dual is denoted by $\C^{\perp_H}$. The code is Hermitian self-orthogonal if $\C\subseteq\C^{\perp_H}$.

\begin{lemma}[\cite{HuffmanPless}]\label{lem:selforth-criterion}
Let $\C$ be a linear code over $\F_q$, where $q\in\{2,3,4\}$.
\begin{enumerate}[label=\textnormal{\arabic*)}]
\item If $q=2$ and the weight of every codeword of $\C$ is divisible by $4$, then $\C$ is Euclidean self-orthogonal.
\item If $q=3$, then $\C$ is Euclidean self-orthogonal if and only if the weight of every codeword of $\C$ is divisible by $3$.
\item If $q=4$, then $\C$ is Hermitian self-orthogonal if and only if every codeword of $\C$ has weight divisible by 2.
\end{enumerate}
\end{lemma}

Finally, we recall the definition of a minimal code and the Ashikhmin–Barg criterion. Let $\C$ be a linear code over $\F_q$. A nonzero codeword $c\in\C$ is minimal if every nonzero $c'\in\C$ with $\supp(c')\subseteq\supp(c)$ satisfies $c'=\lambda c$ for some $\lambda\in\F_q^*$. The code $\C$ is minimal if every nonzero codeword of $\C$ is minimal.

\begin{lemma}[\cite{AshikhminBarg1998}]\label{lem:ashikhmin-barg}
Let $w_{\min}$ and $w_{\max}$ denote the minimum and maximum nonzero Hamming weights of a linear code $\C$ over $\F_q$, respectively. If $\frac{w_{\min}}{w_{\max}}>\frac{q-1}{q}$, then $\C$ is minimal. It is a sufficient condition for minimality.
\end{lemma}

\section{Linear codes from simplicial complexes}\label{sec:construction}

In this section, we present the defining-set construction over $\N_q$ and determine its length and dimension over $\F_q$ for general simplicial complexes. We then derive a general Lee-weight formula and use it to determine the parameters and Lee weight distributions of four classes of codes from complexes with one single maximal element.

\subsection{The generic construction of linear codes}\label{subsec:generic}

Let $D=\{d_1,\ldots,d_n\}\subseteq\N_q^m$ be an ordered finite set. Define the code
\begin{equation}\label{eq:generic-construction}
\C_D=\{c_D(v)=(v\cdot d_1,\ldots,v\cdot d_n):
v\in\N_q^m\},
\end{equation}
where $v\cdot d=\sum_{i=1}^m v_i d_i$ is the inner product over $\N_q$. The set $D$ is called the defining set of $\C_D$. The evaluation map $c_D:\N_q^m\to\N_q^n$ is additive, and $c_D(av)=a c_D(v)$ for $a\in\N_q$, since $\N_q$ is commutative. It is also $\F_q$-linear. Hence, by Definition~\ref{def:ring-code}, the code in Eq.~\eqref{eq:generic-construction} is a linear code over $\N_q$ of length $n=|D|$ and an $\F_q$-subspace of $\N_q^n$. By Proposition~\ref{prop:gray}, its Gray image is a linear code over $\F_q$ of length $3n$.

Choose $D_i\subseteq\F_q^m$, $i=1,2,3$, and write $D=uD_1+u^2D_2+u^3D_3 =\{ut_1+u^2t_2+u^3t_3:t_i\in D_i,\ i=1,2,3\}$. The subscripts distinguish the coefficient sets of $u,u^2,u^3$, respectively. For
$v=u\beta_1+u^2\beta_2+u^3\beta_3$ and $d=ut_1+u^2t_2+u^3t_3$, with $\beta_i,t_i\in\F_q^m$, the coordinate product is given by the following lemma.

\begin{lemma}\label{lem:product}
For $v$ and $d$ as above, $v\cdot d
=
u^2\langle\beta_1,t_1\rangle_q
+u^3\bigl(\langle\beta_1,t_2\rangle_q
+\langle\beta_2,t_1\rangle_q\bigr)$.
In particular, $v\cdot d$ is independent of $\beta_3$ and $t_3$.
\end{lemma}

\begin{proof}
For each coordinate, the product of the two $u$-terms contributes to the coefficient of $u^2$. The products of a $u$-term and a $u^2$-term contribute to the coefficient of $u^3$. Every other product contains $u^4$ or a higher power and is zero. Summing over the $m$ coordinates gives the stated formula.
\end{proof}

Since $\phi(u^2x+u^3y)=(x,y,x+y)$ for $x,y\in\F_q$, we have
\begin{equation}\label{eq:symbol-weight}
\wtL(u^2x+u^3y)=\wtH((x,y,x+y)).
\end{equation}
Thus Lemma~\ref{lem:product} and Eq.~\eqref{eq:symbol-weight} express the Lee weight of a codeword through three Hamming-weight contributions. The independence of $t_3$ also gives the following reduction of the defining set.

\begin{proposition}\label{prop:reduction}
Let $\widehat D=uD_1+u^2D_2+u^3D_3$ with $D_3\ne\varnothing$, and let $D=uD_1+u^2D_2$. Then, up to a permutation of coordinates, $\C_{\widehat D}$ is the $|D_3|$-fold coordinate repetition of $\C_D$. In particular, $|\C_{\widehat D}|=|\C_D|$ and
\[
\wtL(c_{\widehat D}(v))=|D_3|\wtL(c_D(v))
\]
for every $v\in\N_q^m$.
\end{proposition}

\begin{proof}
The expression $ut_1+u^2t_2+u^3t_3$ is unique. For each fixed $(t_1,t_2)\in D_1\times D_2$, Lemma~\ref{lem:product} shows that the coordinate value is the same for all $t_3\in D_3$. Ordering the coordinates by $(t_1,t_2)$ therefore gives $|D_3|$ copies of each coordinate of $c_D(v)$. Repetition preserves the number of codewords, and each repeated coordinate contributes the same Lee weight. The assertions follow.
\end{proof}

\begin{remark}
By Proposition~\ref{prop:reduction}, it suffices to study $D=uD_1+u^2D_2$ and to restore the repetition factor $|D_3|$ when a nonempty third component is present. Taking each of $D_1,D_2$ to be a simplicial complex or its complement gives the defining sets $D=u\Delta_1+u^2\Delta_2$, $D=u\Delta_1^c+u^2\Delta_2$, $D=u\Delta_1+u^2\Delta_2^c$, and $D=u\Delta_1^c+u^2\Delta_2^c$. Here $\Delta_1,\Delta_2$ denote the two general complexes, while $D_1,D_2$ denote the component sets selected from them or their complements.
\end{remark}

We first determine the length and dimension when both components are simplicial complexes. Let $\Delta_1$ and $\Delta_2$ be nonempty complexes whose maximal elements have supports $\mathcal A_1=\{F_1,\ldots,F_{l_1}\}$ and $\mathcal A_2=\{H_1,\ldots,H_{l_2}\}$, respectively. Thus $\Delta_1=\langle\mathcal A_1\rangle$ and $\Delta_2=\langle\mathcal A_2\rangle$.

\begin{proposition}\label{prop:general-dimension}
Let $D=u\Delta_1+u^2\Delta_2$. Then $\C_D$ has length
$|\Delta_1||\Delta_2|$ and dimension
\[
\left|\bigcup_{i=1}^{l_1}F_i\right|
+
\left|
\left(\bigcup_{i=1}^{l_1}F_i\right)
\cup
\left(\bigcup_{j=1}^{l_2}H_j\right)
\right|
\]
over $\F_q$. Hence
\[
|\C_D|
=
q^{
\left|\bigcup_{i=1}^{l_1}F_i\right|
+
\left|
(\bigcup_{i=1}^{l_1}F_i)\cup(\bigcup_{j=1}^{l_2}H_j)
\right|
}.
\]
The Gray image $\phi(\C_D)$ has length $3|\Delta_1||\Delta_2|$ and the same dimension over $\F_q$.
\end{proposition}

\begin{proof}
The decomposition $ut_1+u^2t_2$ is unique, so $|D|=|\Delta_1||\Delta_2|$. We next determine the kernel of $c_D$. By Lemma~\ref{lem:product}, a vector $v=u\beta_1+u^2\beta_2+u^3\beta_3$ belongs to this kernel if and only if $\langle\beta_1,t_1\rangle_q=0$ and $\langle\beta_1,t_2\rangle_q+
\langle\beta_2,t_1\rangle_q=0$ for all $(t_1,t_2)\in\Delta_1\times\Delta_2$. Both complexes contain zero. Taking $t_1=0$ in the second equality gives $\beta_1\in\Delta_2^\perp$, and taking $t_2=0$ gives $\beta_2\in\Delta_1^\perp$. The first equality gives $\beta_1\in\Delta_1^\perp$. Conversely, these orthogonality conditions make both equalities hold. Thus the kernel is described by $\beta_1\in\Delta_1^\perp\cap\Delta_2^\perp$ and $\beta_2\in\Delta_1^\perp$, with $\beta_3$ arbitrary.

The span of a simplicial complex is the coordinate subspace supported on the union of its maximal supports. For $\Delta_1$ and $\Delta_2$, these are the unions of the members of $\mathcal A_1$ and $\mathcal A_2$, respectively. Indeed, every vector of the complex is supported on that union, and every coordinate unit vector indexed by the union belongs to at least one of its coordinate subspaces. Consequently, $\Span_{\F_q}(\Delta_1)=\Delta_{\cup_{i=1}^{l_1}F_i}$ and $\Span_{\F_q}(\Delta_2)
=\Delta_{\cup_{j=1}^{l_2}H_j}$. The kernel description therefore yields
\[
\begin{aligned}
\dim_{\F_q}\ker(c_D)
={}&3m-\left|\bigcup_{i=1}^{l_1}F_i\right|-\left|
\left(\bigcup_{i=1}^{l_1}F_i\right)
\cup
\left(\bigcup_{j=1}^{l_2}H_j\right)
\right|.
\end{aligned}
\]
Rank--nullity now gives the dimension and size of $\C_D$. Since $\phi$ is an $\F_q$-linear bijection, its Gray image has the same dimension as $\C_D$ and has length $3|\Delta_1||\Delta_2|$. This completes the proof.
\end{proof}

When $l_1=l_2=1$, put $F_1=A$ and $H_1=B$. Then $\Delta_1=\Delta_A$ and $\Delta_2=\Delta_B$; the subscripts $A,B$ specify the supports of the single maximal elements. In this case, Proposition~\ref{prop:general-dimension} gives length $q^{|A|+|B|}$ and dimension $|A|+|A\cup B|$ for $\C_D$ when $D=u\Delta_A+u^2\Delta_B$. The four choices of a complex or its complement will be studied in Subsection~\ref{subsec:one-maximal}. We next derive a Lee-weight formula using the character
sums introduced in Subsection~\ref{subsec:useful-lemmas}.

\begin{theorem}\label{thm:master}
Let $D=uD_1+u^2D_2$, where $D_1,D_2\subseteq\F_q^m$ are stable under multiplication by $\F_q^*$. For
$v=u\beta_1+u^2\beta_2+u^3\beta_3$, we have
\begin{equation}\label{eq:master}
\begin{aligned}
\wtL(c_D(v))=\frac{q-1}{q}\Big[&3|D_1||D_2|-|D_2|S_{D_1}(\beta_1)-S_{D_2}(\beta_1)\bigl(S_{D_1}(\beta_2)+S_{D_1}(\beta_1+\beta_2)\bigr)\Big].
\end{aligned}
\end{equation}
\end{theorem}

\begin{proof}
By Lemma~\ref{lem:product}, the Gray image of the coordinate indexed by $(t_1,t_2)\in D_1\times D_2$ is
\[
\begin{aligned}
\phi(v\cdot(ut_1+u^2t_2))=\bigl(
&\langle\beta_1,t_1\rangle_q,\
\langle\beta_1,t_2\rangle_q+\langle\beta_2,t_1\rangle_q,\langle\beta_1,t_2\rangle_q+
\langle\beta_1+\beta_2,t_1\rangle_q\bigr).
\end{aligned}
\]
By Eq.~\eqref{eq:symbol-weight}, the Lee weight counts the nonzero entries in these three components. Applying Lemma~\ref{lem:orth-subspace}, we obtain
\[
\begin{aligned}
\wtL(c_D(v))
={}&3|D_1||D_2|-\frac1q\sum_{\lambda\in\F_q}
\sum_{t_1\in D_1}\sum_{t_2\in D_2}\Big[
\chi(\lambda\langle\beta_1,t_1\rangle_q)\\
&\quad+\chi\bigl(\lambda\langle\beta_1,t_2\rangle_q
+\lambda\langle\beta_2,t_1\rangle_q\bigr)\\
&\quad+\chi\bigl(\lambda\langle\beta_1,t_2\rangle_q
+\lambda\langle\beta_1+\beta_2,t_1\rangle_q\bigr)\Big]\\
={}&\frac{3(q-1)}q|D_1||D_2|
-\frac1q\sum_{\lambda\in\F_q^*}\Big[
|D_2|S_{D_1}(\lambda\beta_1)\\
&\quad+S_{D_2}(\lambda\beta_1)
\bigl(S_{D_1}(\lambda\beta_2)
+S_{D_1}(\lambda(\beta_1+\beta_2))\bigr)\Big].
\end{aligned}
\]
Since both defining sets are stable under nonzero scalar multiplication, each summand in the last expression is independent of $\lambda\in\F_q^*$. This gives Eq.~\eqref{eq:master} and completes the proof.
\end{proof}

\subsection{Four classes of linear codes from simplicial complexes with one single maximal element}\label{subsec:one-maximal}

Throughout this subsection, assume $\varnothing\neq A,B\subsetneq[m]$. Let $\Delta_A,\Delta_B$ be the simplicial complexes with one single maximal element of support $A,B$, respectively. Using Eq.~\eqref{eq:master}, we determine the parameters and Lee weight distributions for the four choices of defining sets. The frequency $A_w$ counts codewords of Lee weight $w$, and is also the Hamming-weight frequency of the Gray image.

For $\Delta_A$ with one single maximal element of support $A\subseteq[m]$, Lemma~\ref{lem:simp-char} gives
\begin{equation}\label{eq:single-sums}
\begin{aligned}
S_{\Delta_A}(y)
&=\begin{cases}
q^{|A|},& \text{if }y\in\Delta_A^\perp;\\
0,& \text{if }y\notin\Delta_A^\perp,
\end{cases}
\text{ and}\quad S_{\Delta_A^c}(y)
&=\begin{cases}
q^m-q^{|A|},& \text{if }y=0;\\
-q^{|A|},& \text{if }y\in\Delta_A^\perp\setminus\{0\};\\
0,& \text{if }y\notin\Delta_A^\perp.
\end{cases}
\end{aligned}
\end{equation}

Notice that $\Delta_A^\perp\cap\Delta_B^\perp=\Delta_{A\cup B}^\perp$. Thus $|\Delta_A^\perp|=q^{m-|A|}$, $|\Delta_B^\perp|=q^{m-|B|}$, and $|\Delta_{A\cup B}^\perp|=q^{m-|A\cup B|}$. By inclusion--exclusion, we also have
\begin{equation}\label{eq:parameter-counts}
\begin{aligned}
\begin{cases}
|\Delta_A^\perp\setminus\Delta_B^\perp|=q^{m-|A|}-q^{m-|A\cup B|},\\
|\Delta_B^\perp\setminus\Delta_A^\perp|=q^{m-|B|}-q^{m-|A\cup B|},\\
|\F_q^m\setminus(\Delta_A^\perp\cup\Delta_B^\perp)|=q^m-q^{m-|A|}-q^{m-|B|}+q^{m-|A\cup B|}.
\end{cases}
\end{aligned}
\end{equation}

The cosets $\Delta_A^\perp$ and $-\beta_1+\Delta_A^\perp$ coincide when $\beta_1\in\Delta_A^\perp$ and are disjoint otherwise. Consequently, Eq.~\eqref{eq:single-sums} gives
\begin{equation}\label{eq:pair-subspace}
\begin{aligned}
&S_{\Delta_A}(\beta_2)+S_{\Delta_A}(\beta_1+\beta_2)=
\begin{cases}
2q^{|A|},& \text{if }\beta_1,\beta_2\in\Delta_A^\perp;\\
q^{|A|},&\text{if }\beta_1\notin\Delta_A^\perp, \beta_2\in\Delta_A^\perp\cup(-\beta_1+\Delta_A^\perp);\\
0,&\text{otherwise}.
\end{cases}
\end{aligned}
\end{equation}
For the complement, the zero arguments must be distinguished from the nonzero elements of $\Delta_A^\perp$. Again by Eq.~\eqref{eq:single-sums},
\begin{equation}\label{eq:pair-complement}
\begin{aligned}
&S_{\Delta_A^c}(\beta_2)+S_{\Delta_A^c}(\beta_1+\beta_2)=
\begin{cases}
2(q^m-q^{|A|}),& \text{if }\beta_1=\beta_2=0;\\
q^m-2q^{|A|},&\text{if }\beta_1\in\Delta_A^\perp\setminus\{0\}, \beta_2\in\{0,-\beta_1\};\\
q^m-q^{|A|},& \text{if }\beta_1\notin\Delta_A^\perp, \beta_2\in\{0,-\beta_1\};\\
-2q^{|A|},& \text{if }\beta_1\in\Delta_A^\perp, \beta_2\in\Delta_A^\perp\setminus\{0,-\beta_1\};\\
-q^{|A|},& \begin{array}[t]{@{}l@{}}
\text{if }\beta_1\notin\Delta_A^\perp,\\
\quad\beta_2\in\bigl(\Delta_A^\perp\cup(-\beta_1+\Delta_A^\perp)\bigr)\setminus\{0,-\beta_1\};
\end{array} \\
0,&\text{otherwise}.
\end{cases}
\end{aligned}
\end{equation}

We calculate $A_w$ by counting the pairs $(\beta_1,\beta_2)\in\F_q^{2m}$ giving Lee weight $w$. Since the evaluation map is $\F_q$-linear, each codeword has as many parameter-pair representations as there are pairs giving the zero codeword. Dividing the pair count by this number gives $A_w$. If listed weight values coincide, their parameter conditions are counted together and the corresponding counts in the table are added.

\subsubsection*{Case I: \texorpdfstring{$D=u\Delta_A+u^2\Delta_B$}{D = u DeltaA + u2 DeltaB}}

\begin{theorem}\label{thm:case1}
Let $D=u\Delta_A+u^2\Delta_B$. Then $\C_D$ has length
$q^{|A|+|B|}$, size $q^{|A|+|A\cup B|}$, and minimum Lee
distance $2(q-1)q^{|A|+|B|-1}$.
The Lee weight distribution of $\C_D$ is listed in the following table.

\begin{center}
\normalfont
\renewcommand{\arraystretch}{1.25}
\begin{tabular}{cc}
\toprule
Lee weight  & Frequency \\
\midrule
$0$ & $1$\\
$2(q-1)q^{|A|+|B|-1}$
&
$q^{|A\cup B|}+2q^{|A\setminus B|}-3$
\\
$3(q-1)q^{|A|+|B|-1}$
&
$q^{|A\cup B|}(q^{|A|}-1)-2q^{|A\setminus B|}+2$
\\
\bottomrule
\end{tabular}
\end{center}
Moreover, $\phi(\C_D)$ is a two-weight
$[3q^{|A|+|B|},\,|A|+|A\cup B|,\,2(q-1)q^{|A|+|B|-1}]$
linear code over $\F_q$.
Its Hamming weight distribution is also given by the table.
\end{theorem}

\begin{proof}
By Proposition~\ref{prop:general-dimension}, the length of $\C_D$ is $|D|=q^{|A|+|B|}$. Substituting
$D_1=\Delta_A$ and $D_2=\Delta_B$ in Eq.~\eqref{eq:master}, we obtain
\begin{equation}\label{eq:case1-calculation}
\begin{aligned}
\wtL(c_D(v))=\frac{q-1}{q}\Big[
&3q^{|A|+|B|}-q^{|B|}S_{\Delta_A}(\beta_1)-S_{\Delta_B}(\beta_1)
\bigl(S_{\Delta_A}(\beta_2)+S_{\Delta_A}(\beta_1+\beta_2)\bigr)
\Big].
\end{aligned}
\end{equation}
By Eqs.~\eqref{eq:single-sums} and \eqref{eq:pair-subspace},
the Lee weight $\wtL(c_D(v))$ is given as follows:
\[
\left\{
\begin{array}{ll}
0, & 
\begin{array}{l}
\text{if } \beta_1\in\Delta_{A\cup B}^\perp, \beta_2\in\Delta_A^\perp;
\end{array} \\[12pt]
w_1:=2(q-1)q^{|A|+|B|-1}, & 
\begin{array}{l}
\text{if } \beta_1\in\Delta_{A\cup B}^\perp, \beta_2\notin\Delta_A^\perp, \\
\quad\text{or } \beta_1\in \Delta_A^\perp\setminus\Delta_B^\perp, \beta_2\in\F_q^m, \\
\quad\text{or } \beta_1\in \Delta_B^\perp\setminus\Delta_A^\perp, \beta_2\in\Delta_A^\perp\cup(-\beta_1+\Delta_A^\perp);
\end{array} \\[12pt]
w_2:=3(q-1)q^{|A|+|B|-1}, & 
\begin{array}{l}
\text{if } \beta_1\in \Delta_B^\perp\setminus\Delta_A^\perp, \beta_2\notin\Delta_A^\perp\cup(-\beta_1+\Delta_A^\perp), \\
\quad\text{or } \beta_1\notin \Delta_A^\perp\cup\Delta_B^\perp, \beta_2\in\F_q^m.
\end{array}
\end{array}
\right.
\]
Notice that $0<w_1<w_2$. Hence the minimum Lee distance is $w_1$. The zero-weight condition gives $q^{m-|A\cup B|}q^{m-|A|}$ pairs $(\beta_1,\beta_2)$. It follows that $|\C_D|=q^{2m}/(q^{m-|A\cup B|}q^{m-|A|}) =q^{|A|+|A\cup B|}$. The zero-weight pairs form $\Delta_{A\cup B}^\perp\times\Delta_A^\perp$. Hence every codeword has $q^{m-|A\cup B|}q^{m-|A|}$ parameter-pair representations.

Now we determine the weight frequencies by counting the parameter pairs satisfying the conditions above. We consider the following cases.

\textit{Case 1: $\beta_1\in\Delta_{A\cup B}^\perp$.}
In this case, $S_{\Delta_A}(\beta_1)=q^{|A|}$ and $S_{\Delta_B}(\beta_1)=q^{|B|}$, and $S_{\Delta_A}(\beta_1+\beta_2)=S_{\Delta_A}(\beta_2)$ because $\beta_1\in\Delta_A^\perp$. Hence
Eq.~\eqref{eq:case1-calculation} becomes
\[
\wtL(c_D(v))=2(q-1)q^{|B|-1}
\bigl(q^{|A|}-S_{\Delta_A}(\beta_2)\bigr).
\]
It is zero for $\beta_2\in\Delta_A^\perp$ and is $w_1$ otherwise. There are $q^{m-|A\cup B|}$ choices of $\beta_1$. For each choice of $\beta_1$, the numbers of choices of $\beta_2$ giving zero and $w_1$ are $q^{m-|A|}$ and $q^m-q^{m-|A|}$, respectively. In particular, there are $q^{m-|A\cup B|}q^{m-|A|}$ zero-weight pairs.

\textit{Case 2: $\beta_1\in\Delta_A^\perp\setminus\Delta_B^\perp$.}
By Eq.~\eqref{eq:single-sums}, we have $S_{\Delta_A}(\beta_1)=q^{|A|}$ and $S_{\Delta_B}(\beta_1)=0$. Therefore, the weight is $\frac{q-1}{q}(3q^{|A|+|B|}-q^{|A|+|B|})=w_1$ for every $\beta_2\in\F_q^m$. By Eq.~\eqref{eq:parameter-counts}, the number of pairs is $(q^{m-|A|}-q^{m-|A\cup B|})q^m$.

\textit{Case 3: $\beta_1\in\Delta_B^\perp\setminus\Delta_A^\perp$.}
In this case, $S_{\Delta_A}(\beta_1)=0$ and $S_{\Delta_B}(\beta_1)=q^{|B|}$, so
\[
\wtL(c_D(v))=(q-1)q^{|B|-1}
\bigl(3q^{|A|}-S_{\Delta_A}(\beta_2)
-S_{\Delta_A}(\beta_1+\beta_2)\bigr).
\]
Since the two cosets are disjoint, the sum in Eq.~\eqref{eq:pair-subspace} is $q^{|A|}$ on their
union and zero outside it. Substitution in Eq.~\eqref{eq:case1-calculation} gives $w_1$ and $w_2$,
respectively. For each of the $q^{m-|B|}-q^{m-|A\cup B|}$ choices of $\beta_1$, the numbers of choices of $\beta_2$ are $2q^{m-|A|}$ and $q^m-2q^{m-|A|}$. Thus, there are $2(q^{m-|B|}-q^{m-|A\cup B|})q^{m-|A|}$ pairs of weight $w_1$ and $(q^{m-|B|}-q^{m-|A\cup B|})(q^m-2q^{m-|A|})$ pairs
of weight $w_2$.

\textit{Case 4: $\beta_1\notin\Delta_A^\perp\cup\Delta_B^\perp$.}
Both character sums at $\beta_1$ vanish, so the weight is $\frac{q-1}{q}3q^{|A|+|B|}=w_2$ for every $\beta_2$. This gives $(q^m-q^{m-|A|}-q^{m-|B|}+q^{m-|A\cup B|})q^m$ pairs.

Adding the pair counts in Cases 1, 2, and 3 and dividing by $q^{m-|A\cup B|}q^{m-|A|}$, we obtain
\[
\begin{aligned}
A_{w_1}
&=\frac{q^{m-|A\cup B|}(q^m-q^{m-|A|})+(q^{m-|A|}-q^{m-|A\cup B|})q^m
+2(q^{m-|B|}-q^{m-|A\cup B|})q^{m-|A|}}{q^{m-|A\cup B|}q^{m-|A|}}\\
&=q^{|A\cup B|}+2q^{|A\setminus B|}-3.
\end{aligned}
\]
Similarly, the pairs giving $w_2$ occur in Cases 3 and 4. After division by the same number of representatives, their total gives
\[
\begin{aligned}
A_{w_2}
&=\frac{(q^{m-|B|}-q^{m-|A\cup B|})(q^m-2q^{m-|A|})
+(q^m-q^{m-|A|}-q^{m-|B|}+q^{m-|A\cup B|})q^m}{q^{m-|A\cup B|}q^{m-|A|}}\\
&=q^{|A\cup B|}(q^{|A|}-1)
-2q^{|A\setminus B|}+2.
\end{aligned}
\]
Both frequencies are positive. Therefore, the Lee weight distribution of $\C_D$ is obtained. By Proposition~\ref{prop:gray}, $\phi(\C_D)$ has length $3|D|$ and the same dimension over $\F_q$ as $\C_D$, and its Hamming weight distribution is the Lee weight distribution above. This completes the proof.
\end{proof}

\begin{example}\label{ex:case1}
Let $q=2$, $m=4$, $A=\{1\}$, and $B=\{2,3\}$, which give $\Delta_A=\{(x_1,0,0,0):x_1\in\F_2\}$ and
$\Delta_B=\{(0,x_2,x_3,0):x_2,x_3\in\F_2\}$. For $D=u\Delta_A+u^2\Delta_B$, Magma computations show that $\phi(\C_D)$ is a $[24,4,8]$ linear code over $\F_2$ with weight enumerator $1+9z^8+6z^{12}$, which is consistent with Theorem~\ref{thm:case1}.
\end{example}

\begin{example}\label{ex:case1.1}
For $q=3$, $m=2$, $A=\{1\}$, and $B=\{2\}$, we have $\Delta_A=\{(x_1,0):x_1\in\F_3\}$ and $\Delta_B=\{(0,x_2):x_2\in\F_3\}$. For $D=u\Delta_A+u^2\Delta_B$, Magma computations show that $\phi(\C_D)$ is a two-weight $[27,3,12]$ linear code over $\F_3$ with weight enumerator $1+12z^{12}+14z^{18}$, which is consistent with Theorem~\ref{thm:case1}.
\end{example}

\subsubsection*{Case II: \texorpdfstring{$D=u\Delta_A^c+u^2\Delta_B$}{D = u DeltaAc + u2 DeltaB}}

\begin{theorem}\label{thm:case2}
Let $D=u\Delta_A^c+u^2\Delta_B$. Then $\C_D$ has length $q^{|B|}(q^m-q^{|A|})$, size $q^{2m}$, and minimum Lee distance $2(q-1)q^{|B|-1}(q^m-q^{|A|})$. The Lee weight distribution of $\C_D$ is listed in the following table.

\begin{center}
\normalfont
\small
\renewcommand{\arraystretch}{1.35}
\begin{tabular}{p{0.40\textwidth}p{0.53\textwidth}}
\toprule
\centering Lee weight & \centering\arraybackslash Frequency \\
\midrule
$0$
&
$1$
\\

$2(q-1)q^{|B|-1}(q^m-q^{|A|})$
&
$q^m-q^{m-|A|}
+2\bigl(q^{m-|B|}-q^{m-|A\cup B|}\bigr)$
\\

$2(q-1)q^{m+|B|-1}$
&
$q^{m-|A|}+2q^{m-|A\cup B|}-3$
\\

$3(q-1)q^{|B|-1}(q^m-q^{|A|})$
&
$\begin{aligned}
&q^m(q^m-q^{m-|A|})\\[-1mm]
&\quad
-2q^{m-|A|}
\bigl(q^{m-|B|}-q^{m-|A\cup B|}\bigr)
\end{aligned}$
\\

$(q-1)q^{|B|-1}(3q^m-2q^{|A|})$
&
$\begin{aligned}
&(q^{m-|A\cup B|}-1)(q^m-q^{m-|A|})\\
&\quad
+2(q^{m-|B|}-q^{m-|A\cup B|})(q^{m-|A|}-1)\\
&\quad
+(q^{m-|A|}-q^{m-|A\cup B|})q^m
\end{aligned}$
\\

$3(q-1)q^{m+|B|-1}$
&
$(q^{m-|A\cup B|}-1)(q^{m-|A|}-2)$
\\
\bottomrule
\end{tabular}
\end{center}
Moreover, $\phi(\C_D)$ is a $[3q^{|B|}(q^m-q^{|A|}),\,2m,\, 2(q-1)q^{|B|-1}(q^m-q^{|A|})]$ linear code over $\F_q$ with at most five nonzero weights. Its Hamming weight distribution is also given by the table.
\end{theorem}

\begin{proof}
By Eq.~\eqref{eq:generic-construction}, the length of $\C_D$ is $|D|$. Since each element of $D$ has a unique expression $ut_1+u^2t_2$, we obtain $|D|=|\Delta_A^c||\Delta_B|=(q^m-q^{|A|})q^{|B|}$.
Substituting $D_1=\Delta_A^c$ and $D_2=\Delta_B$ in Eq.~\eqref{eq:master}, we have
\begin{equation}\label{eq:case2-calculation}
\begin{aligned}
\wtL(c_D(v))=\frac{q-1}{q}\Big[
&3q^{|B|}(q^m-q^{|A|})-q^{|B|}S_{\Delta_A^c}(\beta_1)-S_{\Delta_B}(\beta_1)
\bigl(S_{\Delta_A^c}(\beta_2)+S_{\Delta_A^c}(\beta_1+\beta_2)\bigr)
\Big].
\end{aligned}
\end{equation}
Substituting the values in Eqs.~\eqref{eq:single-sums} and \eqref{eq:pair-complement} into Eq.~\eqref{eq:case2-calculation}, we obtain the following Lee weights:

\[
\left\{
\begin{array}{ll}
0,&
\begin{array}{l}
\text{if }\beta_1=0, \beta_2=0;
\end{array} \\[12pt]
w_1:=2(q-1)q^{|B|-1}(q^m-q^{|A|}),&
\begin{array}{l}
\text{if }\beta_1=0, \beta_2\notin\Delta_A^\perp,\\
\quad\text{or }\beta_1\in \Delta_B^\perp\setminus\Delta_A^\perp, \beta_2\in\{0,-\beta_1\};
\end{array}\\[12pt]
w_2:=2(q-1)q^{m+|B|-1},&
\begin{array}{l}
\text{if }\beta_1=0\text, \beta_2\in\Delta_A^\perp\setminus\{0\},\\
\quad\text{or }\beta_1\in \Delta_{A\cup B}^\perp\setminus\{0\}\text, \beta_2\in\{0,-\beta_1\};
\end{array}\\[12pt]
w_3:=3(q-1)q^{|B|-1}(q^m-q^{|A|}),&
\begin{array}{l}
\text{if }\beta_1\in \Delta_B^\perp\setminus\Delta_A^\perp, \beta_2\notin\Delta_A^\perp\cup(-\beta_1+\Delta_A^\perp),\\
\quad\text{or }\beta_1\notin \Delta_A^\perp\cup\Delta_B^\perp, \beta_2\in\F_q^m;
\end{array}\\[12pt]
w_4:=(q-1)q^{|B|-1}(3q^m-2q^{|A|}),&
\begin{array}{l}
\text{if }\beta_1\in \Delta_{A\cup B}^\perp\setminus\{0\}, \beta_2\notin\Delta_A^\perp,\\
 \quad\text{or }\beta_1\in\Delta_B^\perp\setminus\Delta_A^\perp, \\
 \qquad\beta_2\in\bigl(\Delta_A^\perp\cup(-\beta_1+\Delta_A^\perp)\bigr)\setminus\{0,-\beta_1\}, \\
\quad\text{or }\beta_1\in \Delta_A^\perp\setminus\Delta_B^\perp, \beta_2\in\F_q^m;
\end{array}\\[12pt]
w_5:=3(q-1)q^{m+|B|-1},&
\begin{array}{l}
\text{if }\beta_1\in \Delta_{A\cup B}^\perp\setminus\{0\},\beta_2\in\Delta_A^\perp\setminus\{0,-\beta_1\}.
\end{array}
\end{array}
\right.
\]
Notice that $w_1<w_2$ and $w_1<w_3<w_4<w_5$. Also, $w_1>0$ since $|A|<m$, and it occurs for
$\beta_1=0$ and $\beta_2\notin\Delta_A^\perp$. Therefore the minimum Lee distance is $w_1$. The zero-weight condition is $\beta_1=\beta_2=0$, so $|\C_D|=q^{2m}$. Thus every codeword has a unique parameter pair.

Now we determine the weight frequencies by counting the parameter pairs satisfying the conditions above. We consider the following cases.

\textit{Case 1: $\beta_1=0$.}
Since $S_{\Delta_A^c}(0)=q^m-q^{|A|}$ and $S_{\Delta_B}(0)=q^{|B|}$, we obtain
\[
\wtL(c_D(v))=2(q-1)q^{|B|-1}
\bigl(q^m-q^{|A|}-S_{\Delta_A^c}(\beta_2)\bigr).
\]
The three values of $S_{\Delta_A^c}(\beta_2)$ in Eq.~\eqref{eq:single-sums} give zero when $\beta_2=0$, $w_2$ when $\beta_2\in\Delta_A^\perp\setminus\{0\}$, and $w_1$ when $\beta_2\notin\Delta_A^\perp$. The corresponding numbers of pairs are $1$, $q^{m-|A|}-1$, and $q^m-q^{m-|A|}$, respectively.

\textit{Case 2: $\beta_1\in\Delta_{A\cup B}^\perp\setminus\{0\}$.}
In this case, $S_{\Delta_A^c}(\beta_1)=-q^{|A|}$ and $S_{\Delta_B}(\beta_1)=q^{|B|}$. Thus
\[
\begin{aligned}
\wtL(c_D(v))=(q-1)q^{|B|-1}\Big[
&3q^m-2q^{|A|}-S_{\Delta_A^c}(\beta_2)-S_{\Delta_A^c}(\beta_1+\beta_2)
\Big].
\end{aligned}
\]
If $\beta_2\in\{0,-\beta_1\}$, the sum of the last two character sums is $q^m-2q^{|A|}$, giving $w_2$. If $\beta_2\in\Delta_A^\perp\setminus\{0,-\beta_1\}$, this sum is $-2q^{|A|}$, giving $w_5$.
Outside $\Delta_A^\perp$, both character sums vanish, giving $w_4$. There are $q^{m-|A\cup B|}-1$ choices of $\beta_1$. For each such choice, $0$ and $-\beta_1$ are distinct, so the respective numbers of choices of $\beta_2$ are $2$, $q^{m-|A|}-2$, and $q^m-q^{m-|A|}$. Multiplying by the number of choices of $\beta_1$ gives the three contributions to $w_2$, $w_5$, and $w_4$.

\textit{Case 3: $\beta_1\in\Delta_A^\perp\setminus\Delta_B^\perp$.}
We have $S_{\Delta_A^c}(\beta_1)=-q^{|A|}$ and $S_{\Delta_B}(\beta_1)=0$. Hence the weight is
$(q-1)q^{|B|-1}(3q^m-2q^{|A|})=w_4$ for every $\beta_2$. Consequently, this case contributes $(q^{m-|A|}-q^{m-|A\cup B|})q^m$ parameter pairs.

\textit{Case 4: $\beta_1\in\Delta_B^\perp\setminus\Delta_A^\perp$.}
In this case, $S_{\Delta_A^c}(\beta_1)=0$ and $S_{\Delta_B}(\beta_1)=q^{|B|}$, so
\[
\begin{aligned}
\wtL(c_D(v))=(q-1)q^{|B|-1}\Big[
&3(q^m-q^{|A|})-S_{\Delta_A^c}(\beta_2)-S_{\Delta_A^c}(\beta_1+\beta_2)
\Big].
\end{aligned}
\]
The two cosets in Eq.~\eqref{eq:pair-complement} are disjoint. The sum of the character sums is $q^m-q^{|A|}$ at $\beta_2=0,-\beta_1$, is $-q^{|A|}$ at the other elements of their union, and is zero outside the union. Substitution gives $w_1$, $w_4$, and $w_3$, respectively. There are $q^{m-|B|}-q^{m-|A\cup B|}$ choices of $\beta_1$. For each choice of $\beta_1$, the three numbers of choices of $\beta_2$ are $2$, $2q^{m-|A|}-2$, and $q^m-2q^{m-|A|}$. In particular, the middle count removes the two distinct elements $0,-\beta_1$ from the union.

\textit{Case 5: $\beta_1\notin\Delta_A^\perp\cup\Delta_B^\perp$.}
Both character sums at $\beta_1$ are zero, so the weight is $3(q-1)q^{|B|-1}(q^m-q^{|A|})=w_3$ for every $\beta_2$. By Eq.~\eqref{eq:parameter-counts}, this contributes $(q^m-q^{m-|A|}-q^{m-|B|}+q^{m-|A\cup B|})q^m$ pairs.

We first assume that the listed weights $w_1,\ldots,w_5$ are pairwise distinct. Since every codeword has a unique parameter pair, the counts in Cases 1 and 4 give $A_{w_1}=q^m-q^{m-|A|}+2(q^{m-|B|}-q^{m-|A\cup B|})$. The counts in Cases 1 and 2 give $A_{w_2}=(q^{m-|A|}-1)+2(q^{m-|A\cup B|}-1)=q^{m-|A|}+2q^{m-|A\cup B|}-3$. For $w_3$, Cases 4 and 5 give
\[
\begin{aligned}
A_{w_3}
&=(q^{m-|B|}-q^{m-|A\cup B|})(q^m-2q^{m-|A|})
 +(q^m-q^{m-|A|}-q^{m-|B|}+q^{m-|A\cup B|})q^m\\
&=q^m(q^m-q^{m-|A|})-2q^{m-|A|}(q^{m-|B|}-q^{m-|A\cup B|}).
\end{aligned}
\]
Similarly, adding the counts in Cases 2, 3, and 4 gives
\[
\begin{aligned}
A_{w_4}
={}&(q^{m-|A\cup B|}-1)(q^m-q^{m-|A|})
+2(q^{m-|B|}-q^{m-|A\cup B|})(q^{m-|A|}-1)\\
&+(q^{m-|A|}-q^{m-|A\cup B|})q^m.
\end{aligned}
\]
Finally, Case 2 gives $A_{w_5}=(q^{m-|A\cup B|}-1)(q^{m-|A|}-2)$. The parameter counts in Cases 1--5 remain valid when some listed weights coincide. In that situation, their disjoint parameter conditions are counted together, and the corresponding counts are added to obtain the frequency of the common weight. Thus the table gives the Lee weight distribution for all the stated parameters. By Proposition~\ref{prop:gray}, $\phi(\C_D)$ has length $3|D|$ and the same dimension over $\F_q$ as $\C_D$, and its Hamming weight distribution is the Lee weight distribution above. This completes the proof.
\end{proof}

\begin{corollary}\label{cor:case2-five}
The Gray image in Theorem~\ref{thm:case2} is a five-weight code if and only if $A\cup B\neq[m]$ and
$(q,m-|A|)\notin\{(2,1),(3,1)\}$.
\end{corollary}

\begin{proof}
By the parameter counts in the proof of Theorem~\ref{thm:case2}, the first four counts in the table are positive. The fifth count is $(q^{m-|A\cup B|}-1)(q^{m-|A|}-2)$.
Therefore, all five counts are positive if and only if $A\cup B\neq[m]$ and $q^{m-|A|}>2$. Now we determine when these weights are distinct. Notice that $w_1<w_3<w_4<w_5$ and $w_1<w_2<w_5$. Thus, only $w_2=w_3$ or $w_2=w_4$ can occur. The equality $w_2=w_3$ holds exactly when $q^{m-|A|}=3$, whereas $w_2=w_4$ holds exactly when $q^{m-|A|}=2$. Combining the positivity and distinctness conditions gives the result. This completes the proof.
\end{proof}

\begin{example}\label{ex:case2}
Let $q=2$, $m=4$, $A=\{1\}$, and $B=\{2\}$, which give $\Delta_A=\{(x_1,0,0,0):x_1\in\F_2\}$ and
$\Delta_B=\{(0,x_2,0,0):x_2\in\F_2\}$. For $D=u\Delta_A^c+u^2\Delta_B$, Magma computations show that
$\phi(\C_D)$ is a $[84,8,28]$ linear code over $\F_2$ with weight enumerator $1+16z^{28}+13z^{32}+64z^{42}+144z^{44}+18z^{48}$, which is consistent with Theorem~\ref{thm:case2}.
\end{example}

\begin{example}\label{ex:case2.1}
For $q=3$, $m=3$, $A=\{1\}$, and $B=\{2\}$, we have $\Delta_A=\{(x_1,0,0):x_1\in\F_3\}$ and $\Delta_B=\{(0,x_2,0):x_2\in\F_3\}$. For $D=u\Delta_A^c+u^2\Delta_B$, Magma computations show that $\phi(\C_D)$ is a five-weight $[216,6,96]$ linear code over $\F_3$ with weight enumerator $1+30z^{96}+12z^{108}+378z^{144}+294z^{150}+14z^{162}$, which is consistent with Theorem~\ref{thm:case2} and Corollary~\ref{cor:case2-five}.
\end{example}

\subsubsection*{Case III: \texorpdfstring{$D=u\Delta_A+u^2\Delta_B^c$}{D = u DeltaA + u2 DeltaBc}}

\begin{theorem}\label{thm:case3}
Let $D=u\Delta_A+u^2\Delta_B^c$. Then $\C_D$ has length $q^{|A|}(q^m-q^{|B|})$, size $q^{m+|A|}$, and minimum Lee distance $2(q-1)q^{|A|-1}(q^m-q^{|B|})$. The Lee weight distribution of $\C_D$ is listed in the following table.

\begin{center}
\normalfont
\renewcommand{\arraystretch}{1.3}
\begin{tabular}{cc}
\toprule
Lee weight & Frequency \\
\midrule
$0$
&
$1$
\\
$2(q-1)q^{|A|-1}(q^m-q^{|B|})$
&
$q^m-q^{m-|A\cup B|}$
\\
$2(q-1)q^{m+|A|-1}$
&
$q^{m-|A\cup B|}-1$
\\
$(q-1)q^{|A|-1}(3q^m-2q^{|B|})$
&
$2(q^{m-|B|}-q^{m-|A\cup B|})$
\\
$3(q-1)q^{|A|-1}(q^m-q^{|B|})$
&
$q^m(q^{|A|}-1)
-2(q^{m-|B|}-q^{m-|A\cup B|})$
\\
\bottomrule
\end{tabular}
\end{center}
Moreover, $\phi(\C_D)$ is a $[3q^{|A|}(q^m-q^{|B|}),\,m+|A|,\,2(q-1)q^{|A|-1}(q^m-q^{|B|})]$ linear code over $\F_q$ with at most four nonzero weights. Its Hamming weight distribution is also given by the table.
\end{theorem}

\begin{proof}
By Eq.~\eqref{eq:generic-construction}, the length of $\C_D$ is $|D|$. The expression $ut_1+u^2t_2$ is unique, so $|D|=|\Delta_A||\Delta_B^c|=q^{|A|}(q^m-q^{|B|})$. Taking $D_1=\Delta_A$ and $D_2=\Delta_B^c$ in Eq.~\eqref{eq:master}, we obtain
\begin{equation}\label{eq:case3-calculation}
\begin{aligned}
\wtL(c_D(v))=\frac{q-1}{q}\Big[
&3q^{|A|}(q^m-q^{|B|})-(q^m-q^{|B|})S_{\Delta_A}(\beta_1)\\
&-S_{\Delta_B^c}(\beta_1)
\bigl(S_{\Delta_A}(\beta_2)+S_{\Delta_A}(\beta_1+\beta_2)\bigr)
\Big].
\end{aligned}
\end{equation}
Substituting the values in Eqs.~\eqref{eq:single-sums} and
\eqref{eq:pair-subspace} into Eq.~\eqref{eq:case3-calculation},
we obtain the following Lee weights:

\[
\left\{
\begin{array}{ll}
0,&
\begin{array}{l}
\text{if }\beta_1=0, \beta_2\in\Delta_A^\perp;
\end{array} \\[12pt]
w_1:=2(q-1)q^{|A|-1}(q^m-q^{|B|}),&
\begin{array}{l}
\text{if }\beta_1=0, \beta_2\notin\Delta_A^\perp,\\
\quad\text{or }\beta_1\in \Delta_{A\cup B}^\perp\setminus\{0\}, \beta_2\notin\Delta_A^\perp,\\
\quad\text{or }\beta_1\in \Delta_A^\perp\setminus\Delta_B^\perp, \beta_2\in\F_q^m;
\end{array}\\[12pt]
w_2:=2(q-1)q^{m+|A|-1},&
\begin{array}{l}
\text{if } \beta_1\in \Delta_{A\cup B}^\perp\setminus\{0\}, \beta_2\in\Delta_A^\perp;
\end{array} \\[12pt]
w_3:=(q-1)q^{|A|-1}(3q^m-2q^{|B|}),&
\begin{array}{l}
\text{if } \beta_1\in \Delta_B^\perp\setminus\Delta_A^\perp, \beta_2\in\Delta_A^\perp\cup(-\beta_1+\Delta_A^\perp);
\end{array}\\[12pt]
w_4:=3(q-1)q^{|A|-1}(q^m-q^{|B|}),&
\begin{array}{l}
\text{if }\beta_1\in \Delta_B^\perp\setminus\Delta_A^\perp, \beta_2\notin\Delta_A^\perp\cup(-\beta_1+\Delta_A^\perp),\\
\quad\text{or }\beta_1\notin \Delta_A^\perp\cup\Delta_B^\perp, \beta_2\in\F_q^m.
\end{array}
\end{array}
\right.
\] 
Notice that $w_1<w_2$, $w_1<w_3$, and $w_4=3w_1/2$. Since $|B|<m$, we have $w_1>0$, and $w_1$ occurs for $\beta_1=0$ and $\beta_2\notin\Delta_A^\perp$. Hence the minimum Lee distance is $w_1$. The zero-weight condition is $\beta_1=0$ and $\beta_2\in\Delta_A^\perp$, which gives $q^{m-|A|}$ pairs. It follows that $|\C_D|=q^{2m}/q^{m-|A|}=q^{m+|A|}$. Adding an element of $\Delta_A^\perp$ to $\beta_2$ leaves the codeword and the listed weight conditions unchanged. Thus, every codeword has $q^{m-|A|}$ parameter-pair representations.

Now we determine the weight frequencies by counting the parameter pairs satisfying the conditions above. We consider the following cases.

\textit{Case 1: $\beta_1=0$.}
In this case, $S_{\Delta_A}(0)=q^{|A|}$ and $S_{\Delta_B^c}(0)=q^m-q^{|B|}$, so
\[
\wtL(c_D(v))=\frac{2(q-1)}{q}(q^m-q^{|B|})
\bigl(q^{|A|}-S_{\Delta_A}(\beta_2)\bigr).
\]
It is zero for $\beta_2\in\Delta_A^\perp$ and is $w_1$ otherwise. Thus the respective numbers of pairs are $q^{m-|A|}$ and $q^m-q^{m-|A|}$.

\textit{Case 2: $\beta_1\in\Delta_{A\cup B}^\perp\setminus\{0\}$.}
We have $S_{\Delta_A}(\beta_1)=q^{|A|}$ and $S_{\Delta_B^c}(\beta_1)=-q^{|B|}$. Hence
\[
\begin{aligned}
\wtL(c_D(v))=\frac{q-1}{q}\Big[
&2q^{|A|}(q^m-q^{|B|})+q^{|B|}\bigl(S_{\Delta_A}(\beta_2)
+S_{\Delta_A}(\beta_1+\beta_2)\bigr)\Big].
\end{aligned}
\]
The sum in parentheses is $2q^{|A|}$ for $\beta_2\in\Delta_A^\perp$ and zero otherwise. The resulting weights are $w_2$ and $w_1$, respectively. There are $q^{m-|A\cup B|}-1$ choices of $\beta_1$, and the numbers of choices of $\beta_2$ are $q^{m-|A|}$ and $q^m-q^{m-|A|}$, respectively.

\textit{Case 3: $\beta_1\in\Delta_A^\perp\setminus\Delta_B^\perp$.} Since $S_{\Delta_A}(\beta_1)=q^{|A|}$ and $S_{\Delta_B^c}(\beta_1)=0$, the weight is $2(q-1)q^{|A|-1}(q^m-q^{|B|})=w_1$ for every $\beta_2$. Hence, this case contributes $(q^{m-|A|}-q^{m-|A\cup B|})q^m$ parameter pairs.

\textit{Case 4: $\beta_1\in\Delta_B^\perp\setminus\Delta_A^\perp$.}
In this case, $S_{\Delta_A}(\beta_1)=0$ and $S_{\Delta_B^c}(\beta_1)=-q^{|B|}$. Therefore
\[
\begin{aligned}
\wtL(c_D(v))=\frac{q-1}{q}\Big[
&3q^{|A|}(q^m-q^{|B|})+q^{|B|}\bigl(S_{\Delta_A}(\beta_2)
+S_{\Delta_A}(\beta_1+\beta_2)\bigr)\Big].
\end{aligned}
\]
The sum in parentheses is $q^{|A|}$ on the union of the two disjoint cosets and zero outside it, giving $w_3$ and $w_4$, respectively. There are $q^{m-|B|}-q^{m-|A\cup B|}$ choices of $\beta_1$.
For each choice of $\beta_1$, the two numbers of choices of $\beta_2$ are $2q^{m-|A|}$ and $q^m-2q^{m-|A|}$.

\textit{Case 5: $\beta_1\notin\Delta_A^\perp\cup\Delta_B^\perp$.}
Both character sums at $\beta_1$ vanish, so the weight is $3(q-1)q^{|A|-1}(q^m-q^{|B|})=w_4$ for every $\beta_2$. Hence, this case contributes $(q^m-q^{m-|A|}-q^{m-|B|}+q^{m-|A\cup B|})q^m$ parameter pairs.

We first assume that $w_1,\ldots,w_4$ are pairwise distinct. Each codeword has $q^{m-|A|}$ parameter-pair representations, so the counts in Cases 1, 2, and 3 give
\[
\begin{aligned}
A_{w_1}
&=\frac{q^m-q^{m-|A|}+(q^{m-|A\cup B|}-1)(q^m-q^{m-|A|})
+(q^{m-|A|}-q^{m-|A\cup B|})q^m}{q^{m-|A|}}\\
&=q^m-q^{m-|A\cup B|}.
\end{aligned}
\]
For $w_2$, Case 2 gives $A_{w_2}=\frac{(q^{m-|A\cup B|}-1)q^{m-|A|}}{q^{m-|A|}}=q^{m-|A\cup B|}-1$. For $w_3$, Case 4 gives $A_{w_3}=\frac{2(q^{m-|B|}-q^{m-|A\cup B|})q^{m-|A|}}{q^{m-|A|}}=2(q^{m-|B|}-q^{m-|A\cup B|})$. Finally, Cases 4 and 5 give
\[
\begin{aligned}
A_{w_4}
&=\frac{(q^{m-|B|}-q^{m-|A\cup B|})(q^m-2q^{m-|A|})
+(q^m-q^{m-|A|}-q^{m-|B|}+q^{m-|A\cup B|})q^m}{q^{m-|A|}}\\
&=q^m(q^{|A|}-1)-2(q^{m-|B|}-q^{m-|A\cup B|}).
\end{aligned}
\]
If some listed weights coincide, we add the pair counts for all their parameter conditions and divide the sum by $q^{m-|A|}$. Equivalently, the corresponding counts in the table are added to obtain the frequency of the common weight. This gives the stated Lee weight distribution. By Proposition~\ref{prop:gray}, $\phi(\C_D)$ has length $3|D|$ and the same dimension over $\F_q$ as $\C_D$, and its Hamming weight distribution is the Lee weight distribution above. This completes the proof.
\end{proof}

\begin{corollary}\label{cor:case3-four}
The Gray image in Theorem~\ref{thm:case3} is a four-weight code if and only if $A\cup B\neq[m]$ and $A\not\subseteq B$.
\end{corollary}

\begin{proof}
By the parameter counts in the proof of Theorem~\ref{thm:case3}, the counts for $w_1$ and $w_4$ are positive. The count for $w_2$ is positive if and only if $A\cup B\neq[m]$, and the count for $w_3$ is positive if and only if $A\not\subseteq B$. Hence, both conditions are necessary for a four-weight code. Conversely, assume these two conditions. They imply $m-|B|\ge2$, so $q^{m-|B|}>3$. We then have $w_4-w_2=(q-1)q^{|A|+|B|-1}(q^{m-|B|}-3)>0$. Moreover, $w_3-w_4=(q-1)q^{|A|+|B|-1}>0$. Together with $w_1<w_2$, these inequalities give $w_1<w_2<w_4<w_3$. Thus, all four weights are distinct
and have positive frequencies. This completes the proof.
\end{proof}

\begin{example}\label{ex:case3}
Let $q=2$, $m=4$, $A=\{1\}$, and $B=\{2,3\}$, which give $\Delta_A=\{(x_1,0,0,0):x_1\in\F_2\}$ and
$\Delta_B=\{(0,x_2,x_3,0):x_2,x_3\in\F_2\}$. For $D=u\Delta_A+u^2\Delta_B^c$, Magma computations show that $\phi(\C_D)$ is a $[72,5,24]$ linear code over $\F_2$ with weight enumerator $1+14z^{24}+z^{32}+12z^{36}+4z^{40}$, which is consistent with Theorem~\ref{thm:case3}.
\end{example}

\begin{example}\label{ex:case3.1}
For $q=5$, $m=3$, $A=\{1\}$, and $B=\{2\}$, we have $\Delta_A=\{(x_1,0,0):x_1\in\F_5\}$ and
$\Delta_B=\{(0,x_2,0):x_2\in\F_5\}$. For $D=u\Delta_A+u^2\Delta_B^c$, Magma computations show that $\phi(\C_D)$ is a four-weight $[1800,4,960]$ linear code over $\F_5$ with weight enumerator $1+120z^{960}+4z^{1000}+460z^{1440}+40z^{1460}$, which is consistent with Theorem~\ref{thm:case3} and  Corollary~\ref{cor:case3-four}.
\end{example}

\subsubsection*{Case IV: \texorpdfstring{$D=u\Delta_A^c+u^2\Delta_B^c$}{D = u DeltaAc + u2 DeltaBc}}

\begin{theorem}\label{thm:case4}
Let $D=u\Delta_A^c+u^2\Delta_B^c$. Then $\C_D$ has length $(q^m-q^{|A|})(q^m-q^{|B|})$, size $q^{2m}$, and minimum Lee distance $2(q-1)(q^{m-1}-q^{|A|-1})(q^m-q^{|B|})$. The Lee weight distribution of $\C_D$ is listed in the following table.

\begin{center}
\normalfont
\scriptsize
\renewcommand{\arraystretch}{1.35}
\begin{tabular}{p{0.41\textwidth}p{0.52\textwidth}}
\toprule
\centering Lee weight & \centering\arraybackslash Frequency \\
\midrule
$0$
&
$1$
\\

$2(q-1)q^{m-1}(q^m-q^{|B|})$
&
$q^{m-|A|}-1$
\\

$2(q-1)(q^{m-1}-q^{|A|-1})(q^m-q^{|B|})$
&
$q^m-q^{m-|A|}$
\\

$(q-1)q^{m-1}(3q^m-2q^{|A|}-2q^{|B|})$
&
$2(q^{m-|A\cup B|}-1)$
\\

$(q-1)q^{m-1}(3q^m-2q^{|A|}-3q^{|B|})$
&
$(q^{m-|A\cup B|}-1)(q^{m-|A|}-2)$
\\

$(q-1)(q^{m-1}-q^{|B|-1})(3q^m-2q^{|A|})$
&
$\begin{aligned}
&(q^{m-|A\cup B|}-1)(q^m-q^{m-|A|})\\
&\quad
+(q^{m-|A|}-q^{m-|A\cup B|})q^m
\end{aligned}$
\\

$(q-1)(q^{m-1}-q^{|A|-1})(3q^m-2q^{|B|})$
&
$2(q^{m-|B|}-q^{m-|A\cup B|})$
\\

$\begin{aligned}
&(q-1)\bigl(
3q^{2m-1}-3q^{m+|A|-1}\\[-1mm]
&\qquad
-3q^{m+|B|-1}+2q^{|A|+|B|-1}
\bigr)
\end{aligned}$
&
$2(q^{m-|B|}-q^{m-|A\cup B|})(q^{m-|A|}-1)$
\\

$3(q-1)(q^{m-1}-q^{|A|-1})(q^m-q^{|B|})$
&
$\begin{aligned}
&q^m(q^m-q^{m-|A|})\\[-1mm]
&\quad
-2q^{m-|A|}
(q^{m-|B|}-q^{m-|A\cup B|})
\end{aligned}$
\\
\bottomrule
\end{tabular}
\end{center}
Moreover, $\phi(\C_D)$ is a $[3(q^m-q^{|A|})(q^m-q^{|B|}),\,2m,\, 2(q-1)(q^{m-1}-q^{|A|-1})(q^m-q^{|B|})]$ linear code over $\F_q$ with at most eight nonzero weights. Its Hamming weight distribution is also given by the table.
\end{theorem}

\begin{proof}
By Eq.~\eqref{eq:generic-construction}, the length of $\C_D$ is $|D|$. The uniqueness of the expression $ut_1+u^2t_2$ gives $|D|=|\Delta_A^c||\Delta_B^c|=(q^m-q^{|A|})(q^m-q^{|B|})$.
Taking $D_1=\Delta_A^c$ and $D_2=\Delta_B^c$ in Eq.~\eqref{eq:master}, we have
\begin{equation}\label{eq:case4-calculation}
\begin{aligned}
\wtL(c_D(v))=\frac{q-1}{q}\Big[
&3(q^m-q^{|A|})(q^m-q^{|B|})-(q^m-q^{|B|})S_{\Delta_A^c}(\beta_1)\\
&-S_{\Delta_B^c}(\beta_1)
\bigl(S_{\Delta_A^c}(\beta_2)+S_{\Delta_A^c}(\beta_1+\beta_2)\bigr)
\Big].
\end{aligned}
\end{equation}
Substituting the values in Eqs.~\eqref{eq:single-sums} and
\eqref{eq:pair-complement} into Eq.~\eqref{eq:case4-calculation},
we obtain the following values of $\wtL(c_D(v))$:

\[
\left\{
\begin{array}{ll}
0,&\begin{array}{l}
\text{if }\beta_1=0,\beta_2=0;
\end{array} \\[12pt]
w_1:=2(q-1)q^{m-1}(q^m-q^{|B|}),&\begin{array}{l}
\text{if }\beta_1=0,\\
\quad\beta_2\in\Delta_A^\perp\setminus\{0\};
\end{array}\\[12pt]
w_2:=2(q-1)(q^{m-1}-q^{|A|-1})(q^m-q^{|B|}),
&\begin{array}{l}
\text{if }\beta_1=0,\\
\quad \beta_2\notin\Delta_A^\perp;
\end{array} \\[12pt]
w_3:=(q-1)q^{m-1}(3q^m-2q^{|A|}-2q^{|B|}),
&\begin{array}{l}
\text{if }\beta_1\in \Delta_{A\cup B}^\perp\setminus\{0\},\\
\quad\beta_2\in\{0,-\beta_1\};
\end{array}\\[12pt]
w_4:=(q-1)q^{m-1}(3q^m-2q^{|A|}-3q^{|B|}),
&\begin{array}{l}
\text{if }\beta_1\in \Delta_{A\cup B}^\perp\setminus\{0\},\\
\quad\beta_2\in\Delta_A^\perp\setminus\{0,-\beta_1\};
\end{array}\\[12pt]
w_5:=(q-1)(q^{m-1}-q^{|B|-1})(3q^m-2q^{|A|}),
&\begin{array}{l}
\text{if }\beta_1\in \Delta_{A\cup B}^\perp\setminus\{0\},\\
\quad\beta_2\notin\Delta_A^\perp,\\
\quad\text{or } \beta_1\in \Delta_A^\perp\setminus\Delta_B^\perp,\\
\quad\beta_2\in\F_q^m;
\end{array}\\[12pt]
w_6:=(q-1)(q^{m-1}-q^{|A|-1})(3q^m-2q^{|B|}),
&\begin{array}{l}
\text{if }\beta_1\in \Delta_B^\perp\setminus\Delta_A^\perp,\\
\quad\beta_2\in\{0,-\beta_1\};
\end{array}\\[12pt]
\begin{aligned}
w_7:={}&(q-1)\bigl(3q^{2m-1}-3q^{m+|A|-1}\\
&\quad{}-3q^{m+|B|-1}+2q^{|A|+|B|-1}\bigr),
\end{aligned}
&\begin{array}{l}
\text{if }\beta_1\in \Delta_B^\perp\setminus\Delta_A^\perp,\\
\quad\beta_2\in\Delta_A^\perp\cup(-\beta_1+\Delta_A^\perp),\\
\quad\beta_2\notin\{0,-\beta_1\};
\end{array}\\[12pt]
w_8:=3(q-1)(q^{m-1}-q^{|A|-1})(q^m-q^{|B|}),
&\begin{array}{l}
\text{if }\beta_1\in \Delta_B^\perp\setminus\Delta_A^\perp,\\
\quad\beta_2\notin\Delta_A^\perp\cup(-\beta_1+\Delta_A^\perp),\\
\quad\text{or }\beta_1\notin \Delta_A^\perp\cup\Delta_B^\perp,\\
\quad\beta_2\in\F_q^m.
\end{array}
\end{array}
\right.
\]

We first determine the parameters of $\C_D$. Since $q^{|A|}\le q^{m-1}$ and $q^{|B|}\le q^{m-1}$, we have $w_4-w_2\ge(q-1)q^{2m-3}\bigl(q(q-1)-2\bigr)\ge0$. Moreover, the same bounds give $w_7-w_2=(q-1)q^{m-1}\bigl(q^m-q^{|A|}-q^{|B|}\bigr)\ge0$. Also, $w_3>w_4$, and each of $w_1,w_5,w_6,w_8$ is greater than $w_2$. The weight $w_2$ is positive and occurs for $\beta_1=0$
and $\beta_2\notin\Delta_A^\perp$. Thus the minimum Lee distance is $w_2$. The zero weight occurs only when $\beta_1=\beta_2=0$, and hence $|\C_D|=q^{2m}$. Therefore the parameter pairs represent distinct codewords.

Now we determine the weight frequencies by counting the parameter pairs satisfying the conditions above. We consider the following cases.

\textit{Case 1: $\beta_1=0$.}
The two character sums at zero are $q^m-q^{|A|}$ and $q^m-q^{|B|}$, respectively. Thus
\[
\wtL(c_D(v))=\frac{2(q-1)}{q}(q^m-q^{|B|})
\bigl(q^m-q^{|A|}-S_{\Delta_A^c}(\beta_2)\bigr).
\]
For $\beta_2=0$, for $\beta_2\in\Delta_A^\perp\setminus\{0\}$, and for $\beta_2\notin\Delta_A^\perp$, the character sum is $q^m-q^{|A|}$, $-q^{|A|}$, and zero, respectively. Substitution gives $0$, $w_1$, and $w_2$. The three pair counts are $1$, $q^{m-|A|}-1$, and $q^m-q^{m-|A|}$.

\textit{Case 2: $\beta_1\in\Delta_{A\cup B}^\perp\setminus\{0\}$.}
In this case, $S_{\Delta_A^c}(\beta_1)=-q^{|A|}$ and $S_{\Delta_B^c}(\beta_1)=-q^{|B|}$. Hence
\[
\begin{aligned}
\wtL(c_D(v))=\frac{q-1}{q}\Big[
&(q^m-q^{|B|})(3q^m-2q^{|A|})+q^{|B|}\bigl(S_{\Delta_A^c}(\beta_2)
+S_{\Delta_A^c}(\beta_1+\beta_2)\bigr)\Big].
\end{aligned}
\]
The sum in parentheses is $q^m-2q^{|A|}$ at $\beta_2=0,-\beta_1$, is $-2q^{|A|}$ on $\Delta_A^\perp\setminus\{0,-\beta_1\}$, and is zero outside $\Delta_A^\perp$. For the first substitution, we have $(q^m-q^{|B|})(3q^m-2q^{|A|})+q^{|B|}(q^m-2q^{|A|})=q^m(3q^m-2q^{|A|}-2q^{|B|})$.
For the second substitution, we have $(q^m-q^{|B|})(3q^m-2q^{|A|})-2q^{|A|+|B|}=q^m(3q^m-2q^{|A|}-3q^{|B|})$.

Consequently the three weights are $w_3$, $w_4$, and $w_5$. There are $q^{m-|A\cup B|}-1$ choices of $\beta_1$. For each choice of $\beta_1$, the numbers of choices of $\beta_2$ are $2$, $q^{m-|A|}-2$, and $q^m-q^{m-|A|}$, respectively. The two excluded elements are distinct because $\beta_1\ne0$.

\textit{Case 3: $\beta_1\in\Delta_A^\perp\setminus\Delta_B^\perp$.}
We have $S_{\Delta_A^c}(\beta_1)=-q^{|A|}$ and $S_{\Delta_B^c}(\beta_1)=0$. Therefore the weight is
$(q-1)(q^{m-1}-q^{|B|-1})(3q^m-2q^{|A|})=w_5$ for every $\beta_2$. This gives $(q^{m-|A|}-q^{m-|A\cup B|})q^m$ pairs.

\textit{Case 4: $\beta_1\in\Delta_B^\perp\setminus\Delta_A^\perp$.}
In this case, $S_{\Delta_A^c}(\beta_1)=0$ and $S_{\Delta_B^c}(\beta_1)=-q^{|B|}$, so
\[
\begin{aligned}
\wtL(c_D(v))=\frac{q-1}{q}\Big[
&3(q^m-q^{|A|})(q^m-q^{|B|})+q^{|B|}\bigl(S_{\Delta_A^c}(\beta_2)
+S_{\Delta_A^c}(\beta_1+\beta_2)\bigr)\Big].
\end{aligned}
\]
For $\beta_2=0,-\beta_1$, the sum in parentheses is $q^m-q^{|A|}$, and the expression in brackets becomes $(q^m-q^{|A|})(3q^m-2q^{|B|})$, giving $w_6$. At the other elements of the two disjoint cosets, the sum is $-q^{|A|}$. In this case, $3(q^m-q^{|A|})(q^m-q^{|B|})-q^{|A|+|B|}=3q^{2m}-3q^{m+|A|}-3q^{m+|B|}+2q^{|A|+|B|}$, which gives $w_7$ after multiplication by $(q-1)/q$.
Outside the union, the sum is zero, giving $w_8$. There are $q^{m-|B|}-q^{m-|A\cup B|}$ choices of
$\beta_1$. For each choice of $\beta_1$, the three numbers of choices of $\beta_2$ are $2$, $2q^{m-|A|}-2$, and $q^m-2q^{m-|A|}$, respectively.

\textit{Case 5: $\beta_1\notin\Delta_A^\perp\cup\Delta_B^\perp$.}
Both character sums at $\beta_1$ are zero. Thus the weight is $3(q-1)(q^{m-1}-q^{|A|-1})(q^m-q^{|B|})=w_8$ for every $\beta_2$, giving $(q^m-q^{m-|A|}-q^{m-|B|}+q^{m-|A\cup B|})q^m$ pairs.

We first assume that $w_1,\ldots,w_8$ are pairwise distinct. Since the parameter pairs represent distinct codewords, Case 1 gives
$A_{w_1}=q^{m-|A|}-1$ and $A_{w_2}=q^m-q^{m-|A|}$. The counts in Case 2 give $A_{w_3}=2(q^{m-|A\cup B|}-1)$ and $A_{w_4}=(q^{m-|A\cup B|}-1)(q^{m-|A|}-2)$. For $w_5$, adding the counts in Cases 2 and 3 yields $A_{w_5}=(q^{m-|A\cup B|}-1)(q^m-q^{m-|A|})+(q^{m-|A|}-q^{m-|A\cup B|})q^m$. For $w_6$ and $w_7$, Case 4 gives $A_{w_6}=2(q^{m-|B|}-q^{m-|A\cup B|})$ and $A_{w_7}=2(q^{m-|B|}-q^{m-|A\cup B|})(q^{m-|A|}-1)$. Finally, Cases 4 and 5 give
\[
\begin{aligned}
A_{w_8}
&=(q^{m-|B|}-q^{m-|A\cup B|})(q^m-2q^{m-|A|})
 +(q^m-q^{m-|A|}-q^{m-|B|}+q^{m-|A\cup B|})q^m\\
&=q^m(q^m-q^{m-|A|})-2q^{m-|A|}(q^{m-|B|}-q^{m-|A\cup B|}).
\end{aligned}
\]
If some listed weights coincide, the parameter conditions giving their common value are counted together. Since these conditions are disjoint and each codeword has a unique parameter pair, their counts are added to give the frequency of that weight. The resulting frequencies give the Lee weight distribution in the statement. By Proposition~\ref{prop:gray}, $\phi(\C_D)$ has length $3|D|$ and the same dimension over $\F_q$ as $\C_D$, and its Hamming weight distribution is the Lee weight distribution above. This completes the proof.
\end{proof}

\begin{corollary}\label{cor:case4-eight}
The Gray image in Theorem~\ref{thm:case4} is an eight-weight code if and only if $A\cup B\neq[m]$, $A\not\subseteq B$, $|A|\neq|B|$, and $(q,m-|A|)\notin\{(2,1),(3,1)\}$.
\end{corollary}

\begin{proof}
By the parameter counts in the proof of Theorem~\ref{thm:case4}, all eight counts in the table are positive if and only if $A\cup B\neq[m]$, $A\not\subseteq B$, and $q^{m-|A|}>2$. To determine when these weights are pairwise distinct, we first compute $w_8-w_1=(q-1)q^{|A|-1}(q^m-q^{|B|})(q^{m-|A|}-3)$. Similarly, we have $w_5-w_6=(q-1)q^{m-1}(q^{|A|}-q^{|B|})$. Thus $q^{m-|A|}=3$ gives $w_8=w_1$, and $|A|=|B|$ gives $w_5=w_6$. Therefore the conditions in the statement are necessary.

Conversely, assume those conditions. Then $q^{m-|A|}\ge4$. Also, $A\cup B\neq[m]$ and $A\not\subseteq B$ imply $|B|\le m-2$, so $q^{m-|B|}\ge4$. Hence $q^{|A|}\le q^m/4$ and $q^{|B|}\le q^m/4$. In particular, we obtain $w_7-w_1 =(q-1)q^{m-1}(q^m-3q^{|A|}-q^{|B|})+2(q-1)q^{|A|+|B|-1}>0$. Also, the bound on $q^{|B|}$ gives $w_4-w_8=(q-1)q^{|A|-1}(q^m-3q^{|B|})>0$. The remaining direct comparisons give $w_2<w_1<w_7<w_8<w_4<w_5<w_3$ and $w_8<w_6<w_3$. Thus only $w_4=w_6$ or $w_5=w_6$ could occur. The second equality is excluded by $|A|\neq|B|$. For the first, we have $w_4-w_6=(q-1)q^{|A|+|B|-1} \bigl(q^{m-|B|}-q^{m-|A|}-2\bigr)$. If $q^{m-|B|}>q^{m-|A|}$, their difference is at least $(q-1)q^{m-|A|}\ge4$; otherwise their difference is nonpositive. It therefore cannot equal $2$, and $w_4\neq w_6$. All eight weights are consequently distinct and have positive frequencies. This completes the proof.
\end{proof}

\begin{example}\label{ex:case4}
Let $q=2$, $m=4$, $A=\{1\}$, and $B=\{2,3\}$, which give $\Delta_A=\{(x_1,0,0,0):x_1\in\F_2\}$ and
$\Delta_B=\{(0,x_2,x_3,0):x_2,x_3\in\F_2\}$. For $D=u\Delta_A^c+u^2\Delta_B^c$, Magma computations show that $\phi(\C_D)$ is a $[504,8,168]$ linear code over $\F_2$ with weight enumerator $1+8z^{168}+7z^{192}+28z^{248}+96z^{252}+6z^{256}+104z^{264}+4z^{280}+2z^{288}$, which is consistent with Theorem~\ref{thm:case4}.
\end{example}

\begin{example}\label{ex:case4.1}
For $q=4$, $m=3$, $A=\{1,2\}$, and $B=\{1\}$, we have $\Delta_A=\{(x_1,x_2,0):x_1,x_2\in\F_4\}$ and
$\Delta_B=\{(x_1,0,0):x_1\in\F_4\}$. For $D=u\Delta_A^c+u^2\Delta_B^c$, Magma computations show that
$\phi(\C_D)$ is an eight-weight $[8640,6,4320]$ linear code over $\F_4$ with weight enumerator $1+60z^{4320}+3z^{5760}+72z^{6432}+3744z^{6480}+24z^{6624}+6z^{7104}+180z^{7200}+6z^{7296}$, which is consistent with Theorem~\ref{thm:case4} and Corollary~\ref{cor:case4-eight}.
\end{example}

\begin{remark}
By Theorems~\ref{thm:case1}--\ref{thm:case4}, the Gray images in Cases~I--IV have exactly two, at most five, at most four, and at most eight nonzero weights, respectively. For Cases~II--IV, the conditions for having exactly five, four, and eight nonzero weights are given in Corollaries~\ref{cor:case2-five}, \ref{cor:case3-four}, and~\ref{cor:case4-eight}, respectively. The different spectra obtained by placing a complement in the $u$-component or in the $u^2$-component reflect the asymmetric way in which the mixed product $u\cdot u^2=u^3$ enters the coordinate formula.
\end{remark}

\section[Optimal, self-orthogonal and minimal linear codes]{ Optimal, self-orthogonal and minimal linear codes}\label{sec:applications}

Throughout this section, assume $\varnothing\neq A,B\subsetneq[m]$. We first construct projective codes by puncturing the Gray images in Cases~I--IV and apply the Griesmer bound to study their distance optimality. We then investigate self-orthogonality of the Gray images and give sufficient conditions for minimality of the Gray images and the projective codes.

\subsection[Projective codes]{Projective codes}

For $j\ge0$, put $[j]_q=(q^j-1)/(q-1)$. For $j\ge1$, $[j]_q$ is the number of one-dimensional subspaces of $\F_q^j$. Recall that a linear code is projective if a generator matrix has no zero column and no two proportional columns; equivalently, its dual minimum distance is at least $3$~\cite[Corollary~1.4.14]{HuffmanPless}.

By Lemma~\ref{lem:product}, the image of the map $(\beta_1,\beta_2)\mapsto\phi(c_D(u\beta_1+u^2\beta_2))$ is $\phi(\C_D)$. For $(t_1,t_2)\in D_1\times D_2$, the three corresponding columns of the matrix of this evaluation, written as vectors in $\F_q^{2m}$, are
\begin{equation}\label{eq:three-columns}
\left\{
\begin{aligned}
g_1&=(t_1,0),\\
g_2&=(t_2,t_1),\\
g_3&=(t_1+t_2,t_1).
\end{aligned}
\right.
\end{equation}
Deleting the zero columns and retaining one column from each class of proportional nonzero columns gives $\overline\C$.

\begin{proposition}\label{prop:projective}

In each of the four cases, $\overline\C$ is a projective linear code over $\F_q$ with $\dim_{\F_q}\overline\C=\dim_{\F_q}\phi(\C_D)$ and $d(\overline\C^\perp)=3$.
In Case~I, the dimension is $2|A|+|B|-|A\cap B|$ and the length is
\begin{equation}\label{eq:projective-length-i}
\overline n=2[|A|+|B|]_q-[|B|+|A\cap B|]_q+[|A|]_q-[|A\cap B|]_q.
\end{equation}
In Case~II, the dimension is $2m$ and the length is
\begin{equation}\label{eq:projective-length-ii}
\overline n=(2q^{|B|}+1)([m]_q-[|A|]_q)-[2|B|]_q+[|B|+|A\cap B|]_q.
\end{equation}
In Case~III, the dimension is $m+|A|$ and the length is
\begin{equation}\label{eq:projective-length-iii}
\overline n=[m+|A|]_q-[|B|+|A\cap B|]_q+[|A\cap B|]_q.
\end{equation}
In Case~IV, the dimension is $2m$ and the length is
\[
\overline n
=[2m]_q-[m+|A|]_q+[m]_q-[|A|]_q
-[2|B|]_q+[|B|+|A\cap B|]_q.
\]
\end{proposition}

\begin{proof}
Deleting zero and proportional columns preserves the column span and hence the rank. The retained columns represent distinct projective points, so $d(\overline\C^\perp)\ge3$. In each case, $D_1$ contains a nonzero vector and $D_2$ is nonempty. For any pair with $t_1\ne0$, the three columns in
Eq.~\eqref{eq:three-columns} are nonzero and pairwise nonproportional. Indeed, $g_1$ has zero second block, while any proportionality between $g_2$ and $g_3$ would force the scalar to be $1$ and then $t_1=0$. Since $g_3=g_1+g_2$, their retained representatives are linearly dependent. Hence $d(\overline\C^\perp)=3$.

The dimensions follow from Theorems~\ref{thm:case1}--\ref{thm:case4}. We count the retained projective points in the four cases.

In Case~I, the column sets of types $g_1$, $g_2$, and $g_3$ are the subspaces $\Delta_A\times\{0\}$, $\Delta_B\times\Delta_A$, and $\{(x,y):y\in\Delta_A,\ x-y\in\Delta_B\}$, respectively. The last two have dimension $|A|+|B|$. Their intersection is $\Delta_B\times\Delta_{A\cap B}$, since $x,x-y\in\Delta_B$ if and only if $x,y\in\Delta_B$. Each of them intersects $\Delta_A\times\{0\}$ in $\Delta_{A\cap B}\times\{0\}$, which is also the intersection of all three. Inclusion--exclusion on their projective points gives Eq.~\eqref{eq:projective-length-i}.

In Case~II, the columns with second block $y=0$ are those of type $g_1$, so their first blocks range over $\Delta_A^c$. For $y\in\Delta_A^c$, the first blocks are the union of $\Delta_B$ and $y+\Delta_B$, from $g_2$ and $g_3$, respectively. There are no columns with $y\in\Delta_A\setminus\{0\}$.
The two cosets are disjoint except for the $q^{|B|}-q^{|A\cap B|}$ vectors $y\in\Delta_B\setminus\Delta_A$, for which they coincide. Counting both cosets for each $y\in\Delta_A^c$ and subtracting the overlaps gives a total of $(2q^{|B|}+1)(q^m-q^{|A|})-q^{|B|}(q^{|B|}-q^{|A\cap B|})$ distinct nonzero column vectors, including those with $y=0$. This set of vectors is stable under multiplication by nonzero scalars. Dividing its size by $q-1$ gives Eq.~\eqref{eq:projective-length-ii}.

In Case~III, the columns lie in $\F_q^m\times\Delta_A$ and span this space by Theorem~\ref{thm:case3}. For $y\in\Delta_A\setminus\{0\}$, the point represented by $(x,y)$ is missing precisely when both $x$ and $x-y$ lie in $\Delta_B$, or equivalently when $x,y\in\Delta_B$. This follows by using $t_1=y$ in the columns of types $g_2$ and $g_3$; nonzero scalar multiplication preserves the defining sets. For $y=0$, the point is present if and only if $x\in\Delta_A\setminus\{0\}$ or $x\in\Delta_B^c$, using $g_1$ or using $g_2$ with $t_1=0$,
respectively. Thus the missing points in Case~III are the points of $\Delta_B\times\Delta_{A\cap B}$ other than those of $\Delta_{A\cap B}\times\{0\}$. Subtracting their number from $[m+|A|]_q$ gives Eq.~\eqref{eq:projective-length-iii}.

In Case~IV, we count the points of $\operatorname{PG}(2m-1,q)$ that are not represented by the columns in Eq.~\eqref{eq:three-columns}. First consider points $[(x,y)]$ with $y\in\Delta_A$. If $y\neq0$, such a point cannot be represented by $g_2$ or $g_3$, since the second block of these columns is $t_1\in\Delta_A^c$. If $y=0$, the point is represented by $g_1$ exactly when $x\in\Delta_A^c$. Hence the number of missing points in this case is $[m+|A|]_q-[m]_q+[|A|]_q$. Now let $y\notin\Delta_A$. After projective rescaling, we may take $t_1=y$. The point $[(x,y)]$ is not represented by a column
of type $g_2$ precisely when $x\in\Delta_B$, and it is not represented by a column of type $g_3$ precisely when $x-y\in\Delta_B$. Since $\Delta_B$ is a subspace, both conditions hold if and only if $x,y\in\Delta_B$. Thus the remaining missing points are precisely those with $x\in\Delta_B$ and
$y\in\Delta_B\setminus\Delta_A$, whose number is $[2|B|]_q-[|B|+|A\cap B|]_q$. The two classes of missing points are disjoint. Subtracting their cardinalities from $[2m]_q$ gives the stated length.
\end{proof}

\begin{remark}
In Case~IV, $\phi(\C_D)$ is not projective. Indeed, $|\Delta_B^c|=q^m-q^{|B|}>1$, and hence for fixed $t_1$ the column $g_1=(t_1,0)$ is repeated as $t_2$ varies. Since there are no zero columns, it follows that $d(\phi(\C_D)^\perp)=2$.
\end{remark}

\subsection[Optimal linear codes and examples]{Optimal linear codes and examples}
We first apply the Griesmer bound to the projective codes in Cases~III and~IV of Proposition~\ref{prop:projective} and then conclude with remarks on Cases~I and~II.

\begin{samepage}
\begin{theorem}\label{thm:projective-case3}
Let $\overline\C$ be the projective code from Case~III in Proposition~\ref{prop:projective}. Then $\overline\C$ is an $[\overline n,\,k,\,q^{k-1}-q^{|B|+|A\cap B|-1}]$ linear code over $\F_q$, where $k=m+|A|$ and $\overline n$ is given by Eq.~\eqref{eq:projective-length-iii}. Its weight distribution is given below, with the row containing $q^{|A\cap B|-1}$ omitted when $A\cap B=\varnothing$. Thus $\overline\C$ has two nonzero weights if $A\cap B=\varnothing$ and three otherwise. Moreover, the code is distance-optimal if
\begin{equation}\label{eq:case3-projective-optimality}
[|A\cap B|]_q<|B|+|A\cap B|.
\end{equation}
In particular, it is a Griesmer code if $A\cap B=\varnothing$, and a distance-optimal near Griesmer code if $|A\cap B|=1$.
\end{theorem}

\begin{center}
\small
\renewcommand{\arraystretch}{1.3}
\begin{tabular}{cc}
\toprule
Weight $w$ & Frequency $A_w$\\
\midrule
$0$ & $1$\\
$q^{k-1}-q^{|B|+|A\cap B|-1}$ & $q^{k-|A\cap B|}-q^{k-|B|-|A\cap B|}$\\
$q^{k-1}-q^{|B|+|A\cap B|-1}+q^{|A\cap B|-1}$
& $q^k-q^{k-|A\cap B|}$\\
$q^{k-1}$ & $q^{k-|B|-|A\cap B|}-1$\\
\bottomrule
\end{tabular}
\end{center}
\end{samepage}

\begin{proof}
By Proposition~\ref{prop:projective}, the ambient column space is $\F_q^m\times\Delta_A$. Put $W=\Delta_B\times\Delta_{A\cap B}$ and $T=\Delta_{A\cap B}\times\{0\}$. The missing points are those of $W$ outside $T$. As the retained columns span the ambient space, the linear forms on this space give distinct codewords. Each nonzero form contributes weight $q^{k-1}$ on all projective points of the space. The weight on the retained points is obtained by subtracting its contribution on $W$ and adding that on $T$.

If a form is nonzero on $W$ and zero on $T$, the weight is $q^{k-1}-q^{|B|+|A\cap B|-1}$. The frequency is the difference between the numbers of forms vanishing on $T$ and on $W$, namely $q^{k-|A\cap B|}-q^{k-|B|-|A\cap B|}$. If the form is nonzero on $T$, it is also nonzero on $W$, so its weight is $q^{k-1}-q^{|B|+|A\cap B|-1}+q^{|A\cap B|-1}$. There are $q^k-q^{k-|A\cap B|}$ such forms, and this case is absent when $T=\{0\}$. Finally, the $q^{k-|B|-|A\cap B|}-1$ nonzero forms vanishing on $W$ have weight $q^{k-1}$. Since $1\le |B|+|A\cap B|<k$ and $|B|>0$, the first and last positive weights have positive frequencies, and the positive weights are in increasing order.

We next study the optimality of $\overline\C$. By the Griesmer bound in Eq.~\eqref{eq:griesmer-bound}, we have
\[
\begin{aligned}
g(k,d)
&=\sum_{i=0}^{k-1}
\left\lceil\frac{q^{k-1}-q^{|B|+|A\cap B|-1}}{q^i}\right\rceil\\
&=\sum_{i=0}^{|B|+|A\cap B|-1}
\bigl(q^{k-1-i}-q^{|B|+|A\cap B|-1-i}\bigr)
+\sum_{i=|B|+|A\cap B|}^{k-1}q^{k-1-i}\\
&=[k]_q-[|B|+|A\cap B|]_q.
\end{aligned}
\]
For $0\le i<|B|+|A\cap B|$, the quotient $d/q^i$ is an integer, so replacing $d$ by $d+1$ increases each ceiling by one. For $|B|+|A\cap B|\le i<k$, both ceilings equal $q^{k-1-i}$, since $0\le(q^{|B|+|A\cap B|-1}-1)/q^i<1$. Hence
\[
g(k,d+1)=[k]_q-[|B|+|A\cap B|]_q+|B|+|A\cap B|.
\]
Comparing this with $\overline n$ in Eq.~\eqref{eq:projective-length-iii}, we obtain $g(k,d+1)>\overline n$ under Eq.~\eqref{eq:case3-projective-optimality}. Moreover, $\overline n-g(k,d)=[|A\cap B|]_q$, which is $0$ or $1$ when $|A\cap B|=0$ or $1$, respectively. In both cases Eq.~\eqref{eq:case3-projective-optimality} holds since $|B|\ge1$.
\end{proof}

The following examples illustrate the cases $|A\cap B|=0$ and $1$ in Theorem~\ref{thm:projective-case3}.

\begin{example}\label{ex:optimal-projective-iii}
Let $q=2$, $m=2$, $A=\{1\}$, and $B=\{2\}$ in Case~III. Then $\Delta_A=\{(x_1,0):x_1\in\F_2\}$ and $\Delta_B=\{(0,x_2):x_2\in\F_2\}$. Magma computations show that $\overline\C$ is a $[6,3,3]$ linear code over $\F_2$ with weight enumerator $1+4z^3+3z^4$, which is consistent with Theorem~\ref{thm:projective-case3}. Since $g(3,3)=3+2+1=6$, it is a Griesmer code.
\end{example}

\begin{example}\label{ex:griesmer-projective-iii-ternary} Let $q=3$, $m=2$, $A=\{1\}$, and $B=\{2\}$ in Case~III. Then $\Delta_A=\{(x_1,0):x_1\in\F_3\}$ and $\Delta_B=\{(0,x_2):x_2\in\F_3\}$. Magma computations show that $\overline\C$ is a $[12,3,8]$ linear code over $\F_3$ with weight enumerator $1+18z^8+8z^9$, which is consistent with Theorem~\ref{thm:projective-case3}. Since $g(3,8)=8+3+1=12$, it is a Griesmer code.
\end{example}

\begin{example}\label{ex:near-projective-iii} Let $q=3$, $m=2$, and $A=B=\{1\}$ in Case~III. Then $\Delta_A=\Delta_B=\{(x_1,0):x_1\in\F_3\}$. Magma computations show that $\overline\C$ is a $[10,3,6]$ linear code over $\F_3$ with weight enumerator $1+6z^6+18z^7+2z^9$, which is consistent with Theorem~\ref{thm:projective-case3}. Since $g(3,6)=9$ and $g(3,7)=11>10$, it is a distance-optimal near Griesmer code.
\end{example}

The same Griesmer comparison also gives distance-optimal binary codes whose lengths exceed the bound by three.

\begin{corollary}\label{cor:projective-iii-gap-three} Let $\overline\C$ be the code in Theorem~\ref{thm:projective-case3}. If $q=2$ and $|A\cap B|=2$, then $\overline\C$ is distance-optimal and $\overline n=g(k,d)+3$. In particular, taking $A=B=\{1,2\}$ for $m\ge3$ gives a family of distance-optimal binary codes with parameters $[2^{m+2}-13,\,m+2,\,2^{m+1}-8]$.
\end{corollary}

\begin{proof}
The proof of Theorem~\ref{thm:projective-case3} gives $\overline n-g(k,d)=[2]_2=3$ and $g(k,d+1)-g(k,d)=|B|+2\ge4$. Hence $g(k,d+1)>\overline n$. The stated parameters follow by substituting $q=2$ and $A=B=\{1,2\}$ in Eq.~\eqref{eq:projective-length-iii} and Theorem~\ref{thm:projective-case3}.
\end{proof}

\begin{example}\label{ex:optimal-projective-iii-gap-three} Let $q=2$, $m=4$, and $A=B=\{1,2\}$ in Case~III. Then $\Delta_A=\Delta_B=\{(x_1,x_2,0,0):x_1,x_2\in\F_2\}$. Magma computations show that
$\overline\C$ is a $[51,6,24]$ linear code over $\F_2$ with weight enumerator $1+12z^{24}+48z^{26}+3z^{32}$, which is consistent with Theorem~\ref{thm:projective-case3} and Corollary~\ref{cor:projective-iii-gap-three}. Since $g(6,24)=48<51<52=g(6,25)$, it is distance-optimal.
\end{example}

We now turn to Case~IV. When $B\subseteq A$, the second class of missing points in the proof of Proposition~\ref{prop:projective} is empty, giving the following weight distribution.

\begin{theorem}\label{thm:projective-nested}
Let $\overline\C$ be the projective code from Case~IV in Proposition~\ref{prop:projective}, and assume that $\varnothing\neq B\subseteq A\subsetneq[m]$. Then $\overline\C$ is a four-weight $[\overline n,\,2m,\,q^{2m-1}-q^{m+|A|-1}]$ linear code over $\F_q$, where $\overline n=[2m]_q-[m+|A|]_q+[m]_q-[|A|]_q$. Its weight distribution is listed in the following table. Moreover, $\overline\C$ is distance-optimal if $[m]_q-[|A|]_q<m+|A|$.
\end{theorem}

\begin{center}
\small
\renewcommand{\arraystretch}{1.3}
\begin{tabular}{cc}
\toprule
Weight $w$ & Frequency $A_w$\\
\midrule
$0$ & $1$\\
$q^{2m-1}-q^{m+|A|-1}$
& $q^m-q^{m-|A|}$\\
$q^{2m-1}-q^{m+|A|-1}+q^{m-1}-q^{|A|-1}$
& $q^m(q^m-q^{m-|A|})$\\
$q^{2m-1}-q^{m+|A|-1}+q^{m-1}$
& $q^m(q^{m-|A|}-1)$\\
$q^{2m-1}$
& $q^{m-|A|}-1$\\
\bottomrule
\end{tabular}
\end{center}

\begin{proof}
The length and dimension follow from Proposition~\ref{prop:projective}, since $A\cap B=B$. We first calculate the weights. A nonzero linear functional on a $j$-dimensional subspace is nonzero on $[j]_q-[j-1]_q=q^{j-1}$ projective points. If its restriction to that subspace is zero, it contributes no nonzero coordinate.

By the proof of Proposition~\ref{prop:projective}, the missing projective points are those represented by $(x,y)$ with $y\in\Delta_A\setminus\{0\}$ and $x\in\F_q^m$, together with those represented by $(x,0)$ with $x\in\Delta_A\setminus\{0\}$. Thus their contribution to a weight is obtained from the contribution of $\F_q^m\times\Delta_A$, by subtracting that of $\F_q^m\times\{0\}$ and adding that of $\Delta_A\times\{0\}$.

For $(\beta_1,\beta_2)\in\F_q^m\times\F_q^m$, let $w$ denote the weight of the codeword given by the linear functional $(x,y)\mapsto\langle\beta_1,x\rangle_q+ \langle\beta_2,y\rangle_q$ on the retained columns. If $\beta_1=0$, the restrictions to $\F_q^m\times\{0\}$ and $\Delta_A\times\{0\}$ are zero. The restriction to $\F_q^m\times\Delta_A$ is nonzero exactly when $\beta_2\notin\Delta_A^\perp$. If $\beta_1\neq0$, the restrictions to $\F_q^m\times\Delta_A$ and $\F_q^m\times\{0\}$ are both nonzero, whereas the restriction to $\Delta_A\times\{0\}$ is nonzero exactly when $\beta_1\notin\Delta_A^\perp$. Subtracting the missing contributions from that of $\operatorname{PG}(2m-1,q)$ gives
\[
w=\begin{cases}
0,&\text{if }\beta_1=\beta_2=0;\\[2pt]
w_1:=q^{2m-1}-q^{m+|A|-1},
&\text{if }\beta_1=0\text{ and }\beta_2\notin\Delta_A^\perp;\\[2pt]
w_2:=q^{2m-1}-q^{m+|A|-1}+q^{m-1}-q^{|A|-1},
&\text{if }\beta_1\notin\Delta_A^\perp;\\[2pt]
w_3:=q^{2m-1}-q^{m+|A|-1}+q^{m-1},
&\text{if }\beta_1\in\Delta_A^\perp\setminus\{0\};\\[2pt]
w_4:=q^{2m-1},
&\text{if }\beta_1=0\text{ and }\beta_2\in\Delta_A^\perp\setminus\{0\}.
\end{cases}
\]
Since $0<|A|<m$, one has $0<w_1<w_2<w_3<w_4$. Consequently, the minimum distance is $d=w_1=q^{2m-1}-q^{m+|A|-1}$.

We next determine the frequencies. The dimension of $\overline\C$ is $2m$, so every parameter pair $(\beta_1,\beta_2)$ gives a distinct codeword. For $w_1$, there are $q^m-q^{m-|A|}$ choices of $\beta_2$ with $\beta_1=0$. For $w_2$, there are $q^m-q^{m-|A|}$ choices of $\beta_1$ and $q^m$ choices of $\beta_2$. For $w_3$, there are $q^{m-|A|}-1$ nonzero choices of $\beta_1\in\Delta_A^\perp$ and $q^m$ choices of $\beta_2$. Finally, $w_4$ is obtained from $\beta_1=0$ and the $q^{m-|A|}-1$ nonzero choices of $\beta_2\in\Delta_A^\perp$. These counts give the stated weight distribution.

Now we study the optimality of $\overline\C$. By Eq.~\eqref{eq:griesmer-bound}, a direct computation gives
\begin{equation}\label{eq:projective-griesmer}
\begin{aligned}
g(2m,d)
&=\sum_{i=0}^{2m-1}
\left\lceil\frac{q^{2m-1}-q^{m+|A|-1}}{q^i}\right\rceil\\
&=\sum_{i=0}^{m+|A|-1}
\bigl(q^{2m-1-i}-q^{m+|A|-1-i}\bigr)
+\sum_{i=m+|A|}^{2m-1}q^{2m-1-i}\\
&=[2m]_q-[m+|A|]_q.
\end{aligned}
\end{equation}
To calculate $g(2m,d+1)$, consider the two sums in Eq.~\eqref{eq:projective-griesmer}. For $0\le i\le m+|A|-1$, the quotient $d/q^i$ is an integer, so replacing $d$ by $d+1$ increases each ceiling by one. For $m+|A|\le i\le2m-1$, both ceilings equal $q^{2m-1-i}$, since $0\le(q^{m+|A|-1}-1)/q^i<1$. It follows that
\begin{equation}\label{eq:projective-griesmer-next}
g(2m,d+1)=[2m]_q-[m+|A|]_q+m+|A|.
\end{equation}
Comparing the length with Eq.~\eqref{eq:projective-griesmer-next}, we obtain $g(2m,d+1)>\overline n$ whenever $[m]_q-[|A|]_q<m+|A|$. Hence $\overline\C$ is distance-optimal under this condition.
\end{proof}

\begin{remark}\label{rem:optimal-projective-iv}
For $B\subseteq A$, the sufficient condition in Theorem~\ref{thm:projective-nested} holds precisely for $q=2$, $m\in\{2,3\}$, and $|A|=m-1$.
\end{remark}
Taking $A=B=[m-1]$ gives the following two examples in Case~IV.

\begin{example}\label{ex:optimal-projective-iv-10}
Let $q=2$, $m=2$, and $A=B=\{1\}$ in Case~IV. Then $\Delta_A=\Delta_B=\{(x_1,0):x_1\in\F_2\}$. Magma computations show that $\overline\C$ is a $[10,4,4]$ linear code over $\F_2$ with weight enumerator $1+2z^4+8z^5+4z^6+z^8$, which is consistent with Theorem~\ref{thm:projective-nested}. Since $g(4,4)=8<10<11=g(4,5)$, it is distance-optimal.
\end{example}

\begin{example}\label{ex:optimal-projective-iv-36}
Let $q=2$, $m=3$, and $A=B=\{1,2\}$ in Case~IV. Then $\Delta_A=\Delta_B=\{(x_1,x_2,0):x_1,x_2\in\F_2\}$. Magma computations show that $\overline\C$ is a $[36,6,16]$ linear code over $\F_2$ with weight enumerator $1+6z^{16}+48z^{18}+8z^{20}+z^{32}$, which is consistent with Theorem~\ref{thm:projective-nested}. Since $g(6,16)=32<36<37=g(6,17)$, it is distance-optimal.
\end{example}

We conclude this subsection with remarks on Cases~I and~II.

\begin{remark}\label{rem:optimal-projective-i}
When $A\subseteq B$ in Case~I, the retained columns in Proposition~\ref{prop:projective} represent all projective points of $\Delta_B\times\Delta_A$. It follows that $\overline\C$ is a $q$-ary simplex code with parameters $[[k]_q,\,k,\,q^{k-1}]$, where $k=|A|+|B|$. Since $g(k,q^{k-1})=[k]_q$, it is a Griesmer code.
\end{remark}

\begin{example}\label{ex:optimal-projective-i}
Let $q=2$, $m=3$, $A=\{1,2\}$, and $B=\{1\}$ in Case~I. Then $A\nsubseteq B$, $\Delta_A=\{(x_1,x_2,0):x_1,x_2\in\F_2\}$, and $\Delta_B=\{(x_1,0,0):x_1\in\F_2\}$. Magma computations show that
$\overline\C$ is a $[13,4,6]$ linear code over $\F_2$ with weight enumerator $1+4z^6+8z^7+3z^8$, which is consistent with the length and dimension in Proposition~\ref{prop:projective}. Since $g(4,6)=12$ and $g(4,7)=14>13$, it is a distance-optimal near Griesmer code.
\end{example}

\begin{remark}\label{rem:optimal-projective-ii}
When $q=2$ and $A=B=[m-1]$, translation by any $t_1\notin\Delta_A$ interchanges $\Delta_A$ and its complement. Thus Eq.~\eqref{eq:three-columns} gives the same column sets in Cases~II and~IV, with the types $g_2$ and $g_3$ exchanged and type $g_1$ unchanged. Hence the binary codes in Examples~\ref{ex:optimal-projective-iv-10} and~\ref{ex:optimal-projective-iv-36} also arise in Case~II.
\end{remark}

\subsection{Self-orthogonal codes}

We apply Lemma~\ref{lem:selforth-criterion} to the Gray images in Cases~I--IV by showing that the Hamming weight of every codeword is divisible by $(q-1)q^{|A|+|B|-1}$.

\begin{theorem}\label{thm:selforth-case4}
Let $\C_D$ be a code in one of the four cases in Subsection~\ref{subsec:one-maximal}.
\begin{enumerate}[label=\textnormal{\arabic*)}]
\item If $q=2$ and $|A|+|B|\ge3$, then $\phi(\C_D)$ is Euclidean self-orthogonal.
\item If $q=3$, then $\phi(\C_D)$ is Euclidean self-orthogonal.
\item If $q=4$, then $\phi(\C_D)$ is Hermitian self-orthogonal.
\end{enumerate}
\end{theorem}

\begin{proof}
By Eqs.~\eqref{eq:single-sums} and~\eqref{eq:master}, both $|D_1|$ and $S_{D_1}(\beta)$ are multiples of $q^{|A|}$ for every $\beta\in\F_q^m$, whether $D_1=\Delta_A$ or $D_1=\Delta_A^c$. Similarly, $|D_2|$ and $S_{D_2}(\beta)$ are multiples of $q^{|B|}$. Thus every weight of $\phi(\C_D)$ is divisible by $(q-1)q^{|A|+|B|-1}$. This is a multiple of $4$ if $q=2$ and $|A|+|B|\ge3$, a multiple of $3$ if $q=3$, and even if $q=4$, since $|A|+|B|\ge2$. The result follows from Lemma~\ref{lem:selforth-criterion}.
\end{proof}

\begin{corollary}\label{cor:selforth-eight}
Assume the conditions of Corollary~\ref{cor:case4-eight}. Then the eight-weight code $\phi(\C_D)$ is Euclidean self-orthogonal for $q=2,3$, and Hermitian self-orthogonal for $q=4$.
\end{corollary}

\begin{proof}
For $q=2$, the assumptions $A,B\neq\varnothing$ and $|A|\neq|B|$ give $|A|+|B|\ge3$. The result follows from Theorem~\ref{thm:selforth-case4}, and the cases $q=3,4$ are immediate.
\end{proof}

\subsection{Minimal linear codes}\label{subsec:minimality}

The minimality of the Gray images and their projective punctured codes is investigated using Lemma~\ref{lem:ashikhmin-barg}.

\begin{theorem}\label{thm:minimality}
Let $\C_D$ be a code in Subsection~\ref{subsec:one-maximal}, where $\varnothing\ne A,B\subsetneq[m]$, and let $\overline\C$ be its projective punctured code in Proposition~\ref{prop:projective}. Then $\phi(\C_D)$ and $\overline\C$ are minimal in each of the following cases.

\begin{enumerate}[label=\textnormal{\arabic*)},leftmargin=*,nosep]
\item In Case~I, $q=2$ or $A\subseteq B$.

\item In Case~II, $q=2$ and either $m-|A|\ge3$ or both $m-|A|=2$ and $A\cup B=[m]$.

\item In Case~III, $m-|B|\ge2$ or $A\nsubseteq B$.

\item In Case~IV, $q^{m+|A|-1}+q^{2|B|-1}<q^{2m-2}$.
\end{enumerate}
\end{theorem}

\begin{proof}
Deleting zero columns and retaining one column from each class of proportional nonzero columns preserves support inclusion in both directions, since the coordinates corresponding to columns in the same class are either all zero or all nonzero in every codeword. By Proposition~\ref{prop:projective}, the puncturing map is injective. Since it is linear, two codewords are proportional if and only if
their punctured images are proportional. Thus $\phi(\C_D)$ is minimal if and only if $\overline\C$ is minimal. In each case, we apply Lemma~\ref{lem:ashikhmin-barg} to one of the two codes.

\noindent\emph{1) Case I.}
By Theorem~\ref{thm:case1}, if $q=2$, then the weight ratio of $\phi(\C_D)$ is $w_{\min}/w_{\max}=2/3>1/2$. If $A\subseteq B$, the simplex code in Remark~\ref{rem:optimal-projective-i} has only one nonzero weight, so $w_{\min}/w_{\max}=1>(q-1)/q$.

\noindent\emph{2) Case II.}
Let $w_1,\ldots,w_5$ be the weights in the proof of Theorem~\ref{thm:case2}. Then $w_{\min}=w_1$ for $\phi(\C_D)$. Suppose that $q=2$. If $A\cup B\ne[m]$, the hypothesis gives $m-|A|\ge3$. In this case, $w_5$ has positive frequency and $w_{\max}=w_5$. Hence
\[
\frac{w_{\min}}{w_{\max}}
=\frac{w_1}{w_5}
=\frac{2}{3}\bigl(1-2^{|A|-m}\bigr)
\ge\frac{7}{12}>\frac12.
\]
If $A\cup B=[m]$, then $w_5$ has zero frequency. The $w_4$ row includes the positive term $(2^{m-|A|}-1)2^m$. Since $m-|A|\ge2$, we have $2^m>2^{|A|+1}$ and hence $w_4>w_2$. As $w_1<w_3<w_4$, this gives $w_{\max}=w_4$ and
\[
\frac{w_{\min}}{w_{\max}}
=\frac{w_1}{w_4}
=\frac{2(1-2^{|A|-m})}{3-2\cdot 2^{|A|-m}}
\ge\frac35>\frac12.
\]

\noindent\emph{3) Case III.}
If $m-|B|\ge2$ or $A\nsubseteq B$, then $m-|B|+|A|-|A\cap B|\ge2$. Thus, by Theorem~\ref{thm:projective-case3}, the weight ratio of $\overline\C$ satisfies
\[
\frac{w_{\min}}{w_{\max}}
=1-q^{-(m-|B|+|A|-|A\cap B|)}
\ge1-q^{-2}>\frac{q-1}{q}.
\]

\noindent\emph{4) Case IV.}
By the proof of Proposition~\ref{prop:projective}, the retained point set for $\overline\C$ is obtained from that for $B=A$ by deleting the points represented by $(x,y)$ with $x\in\Delta_B$ and $y\in\Delta_B\setminus\Delta_A$. A linear form is nonzero on at most $q^{2|B|-1}$ projective points of $\Delta_B\times\Delta_B$. It follows from Theorem~\ref{thm:projective-nested} that $w_{\min}\ge q^{2m-1}-q^{m+|A|-1}-q^{2|B|-1}$ and $w_{\max}\le q^{2m-1}$ for $\overline\C$. Therefore, under the stated condition,
\[
\frac{w_{\min}}{w_{\max}}
\ge 1-\frac{q^{m+|A|-1}+q^{2|B|-1}}{q^{2m-1}}
>\frac{q-1}{q}.
\]
This completes the proof.
\end{proof}

\begin{remark}
Let $q=2$, $A=B=\{1,2\}$, and $m\ge4$ in Case~III. Then $\overline\C$ is distance-optimal by Corollary~\ref{cor:projective-iii-gap-three}. It is also minimal since the proof of Theorem~\ref{thm:minimality} gives $w_{\min}/w_{\max}=1-2^{2-m}>1/2$.
\end{remark}

In what follows, we give some examples of minimal codes among the distance-optimal codes in Remark~\ref{rem:optimal-projective-i} and Theorem~\ref{thm:projective-case3}.

\begin{example}\label{ex:minimal-examples}
Some minimal distance-optimal codes are given as follows:
\begin{enumerate}[label=\textnormal{\arabic*)},leftmargin=*]
\item Let $q=3$, $m=2$, and $A=B=\{1\}$. Magma computations show that the projective code $\overline\C$ in Remark~\ref{rem:optimal-projective-i} is a $[4,2,3]$ linear code over $\F_3$ with weight enumerator $1+8z^3$. By Lemma~\ref{lem:ashikhmin-barg}, $\overline\C$ is minimal since $3/3>2/3$, and it is distance-optimal by the Griesmer bound, with $g(2,3)=4$. This is consistent with Remark~\ref{rem:optimal-projective-i} and Theorem~\ref{thm:minimality}.

\item Let $q=2$, $m=5$, and $A=B=\{1,2\}$. Magma computations show that the projective code $\overline\C$ in Theorem~\ref{thm:projective-case3} is a $[115,7,56]$ linear code over $\F_2$ with weight enumerator $1+24z^{56}+96z^{58}+7z^{64}$. By Lemma~\ref{lem:ashikhmin-barg}, $\overline\C$ is minimal since $56/64>1/2$, and it is distance-optimal by the Griesmer bound since $g(7,57)=116>115$. This is consistent with Theorem~\ref{thm:projective-case3}, Corollary~\ref{cor:projective-iii-gap-three}, and Theorem~\ref{thm:minimality}.

\item The Griesmer code in Example~\ref{ex:griesmer-projective-iii-ternary} is minimal by Lemma~\ref{lem:ashikhmin-barg}, consistent with Theorem~\ref{thm:minimality}.

\item The distance-optimal code in Example~\ref{ex:optimal-projective-iii-gap-three} is minimal by Lemma~\ref{lem:ashikhmin-barg}, consistent with Theorem~\ref{thm:minimality}.
\end{enumerate}
\end{example}

\section{Concluding remarks}\label{sec:conclusion}

In this paper, we constructed four classes of linear codes over $\N_q$ from simplicial complexes of $\F_q^m$ and their complements via the defining-set approach. When each simplicial complex has a single maximal element whose support is a nonempty proper subset of $[m]$, we determined the parameters and Lee weight distributions by evaluating exponential sums. The four classes have two, at most five, at most four, and at most eight nonzero Lee weights, respectively. We also obtained necessary and sufficient conditions for the five-, four-, and eight-weight cases. The Lee weight distribution of each code coincides with the Hamming weight distribution of its Gray image. We then punctured the Gray images to obtain projective codes with dual minimum distance three. We determined their lengths and dimensions in all four cases and their weight distributions in Case~III and in Case~IV when $B\subseteq A$. Using these parameters, we applied the Griesmer bound to establish distance optimality for several families in Case~III, including a binary family that is also minimal for $m\ge4$. The same
bound shows that the binary $[13,4,6]$ code from Case~I and the $[10,4,4]$ and $[36,6,16]$ codes common to Cases~II and~IV are also distance-optimal. The Gray images are Euclidean self-orthogonal for $q=3$ and for $q=2$ when $|A|+|B|\ge3$, and Hermitian self-orthogonal for $q=4$. Finally, we applied the Ashikhmin--Barg criterion to obtain sufficient conditions for minimality of the Gray images and their projective punctured codes.

\end{document}